\documentclass[Journal,letterpaper,NoLineNumbers]{ascelike-new}
\usepackage[utf8]{inputenc}
\usepackage[T1]{fontenc}
\usepackage{lmodern}
\usepackage[hyperfootnotes=false]{hyperref}
\hypersetup{
 colorlinks=true,
 citecolor=black,
 linkcolor=black,
 urlcolor=black}
\usepackage{graphicx}
\usepackage[figurename=Fig.,labelfont=bf,labelsep=period]{caption}
\usepackage{subcaption}
\usepackage{amsmath}
\usepackage{newtxtext,newtxmath}

\usepackage{multirow}
\NameTag{Ghosh, \today}
\begin{document}

\title{Bridge Modal Identification using Single Moving Sensor under Random Traffic Loading}

\author[1]{Dhiraj Ghosh}
\author[2]{Suparno Mukhopadhyay, A.M.ASCE}
\author[3]{ Shaily Jain}

\affil[1]{PhD Candidate, Department of Civil Engineering, Indian Institute of Technology Kanpur, Kanpur, Uttar Pradesh 208016, India. ORCID: https://orcid.org/0009-0003-8702-4610. Email: dghosh@iitk.ac.in}
\affil[2]{Assistant Professor, Department of Civil Engineering, Indian Institute of Technology Kanpur, Kanpur, Uttar Pradesh 208016, India (corresponding author). Email: suparno@iitk.ac.in}

\affil[3]{M.Tech. Student, Department of Civil Engineering, Indian Institute of Technology Kanpur, Kanpur, Uttar Pradesh 208016, India. Email: shaily06081995@gmail.com. \\
Currently: L\&T Energy and Hydrocarbon, Vadodara, Gujarat 390019, India}

\maketitle

\begin{abstract}
Modal identification with high spatial resolution is often essential for assessing the structural integrity of bridges. Traditional approaches to get the mode shapes with high spatial resolution rely on a wide array of fixed sensors, which can lead to high installation and data processing costs. This paper explores the feasibility of utilizing the response recorded by a single moving sensor to identify the modal parameters of a bridge system under different loading conditions, such as known excitation and unknown random traffic-induced vibrations. The sensor traverses the bridge and captures its dynamic response (acceleration). The natural frequencies and damping ratios are identified using the moving sensor data in the frequency domain. In the case of known inputs, these parameters are then used to obtain the mode shapes, expressed as a linear combination of basic orthonormal polynomials (BOPs), with the coefficients of the BOPs in the linear combinations obtained via optimization. A statistical formulation is proposed to estimate the mode shapes in the case of unknown random traffic-induced vibrations, including the effect of road roughness. It is shown that the absolute value of the mode shapes are proportional to the ensemble standard deviation (SD) of the modal responses. This approach requires the sensor to traverse the bridge multiple times, with the mode shapes identified in both the time domain using variances, and in frequency domain through the evolutionary power spectrum of these responses. The random traffic loading is modeled such that vehicle arrival times follow a Poisson distribution, while the mass and velocity of the vehicles are assumed to follow uniform distributions. To incorporate the effect of road roughness, modeled as a homogeneous random field, a vehicle-bridge-interaction (VBI) model is utilized. Numerical validation under the different loading conditions demonstrates that a single moving sensor can be used to identify the modal parameters quite accurately, with high spatial resolution of the identified mode shapes, offering a cost-effective and efficient alternative for bridge health monitoring.
\end{abstract}

\section{Introduction}
In structural monitoring, bridges are typically assessed using two primary approaches: direct and indirect. The direct approach involves installing sensors directly on the bridge~\cite{alhamaydeh2022structural,tokognon2017structural}, which, while effective, can be costly, time-consuming, and logistically challenging, especially on bridges under heavy traffic. 
Even with the advent of wireless sensors and software solutions aimed at streamlining the process, challenges remain, particularly in achieving sufficient spatial resolution for detailed structural assessments~\cite{zeng2023automation,pasca2022pyoma}. To get sufficient spatial resolution in identified mode shapes, a dense array of sensors are required.
However, in the indirect approach, the dynamic properties of bridge structures are extracted from the dynamic response of a passing vehicle equipped with sensors, most commonly accelerometers mounted on its axles.
This method effectively turns the vehicle into a moving sensor, offering a practical solution to some of the limitations faced by the direct approach.  
The emerging paradigm of mobile sensing or moving sensor(s) offers a promising alternative. This approach leverages mobile onboard sensors to capture structural vibrations, identify structural parameters, and assess structural condition without the need for fixed sensors. Mobile sensing has the potential to overcome the limitations of sparse spatial measurements associated with fixed sensing, offering a more flexible and comprehensive solution for structural monitoring~\cite{malekjafarian2022review,matarazzo2022crowdsourcing,mcgetrick2009theoretical,eshkevari2020bridge,yi2021damage,zhang2022detecting}.
Further, mobile sensing can save both time and money by utilizing the same instrumented vehicle to scan multiple bridges.

The concept of extracting bridge frequencies from the dynamic response of a moving vehicle was introduced by Yang et al.~\cite{yang2004extracting,yang2005vehicle}, who modeled the vehicle as a spring mass system and the bridge as a simply supported beam vibrating in its first mode. They found that higher vehicle speeds improve the visibility of bridge frequencies by inducing higher amplitude responses. 
In a subsequent study, Lin and Yang~\cite{lin2005use} confirmed the feasibility of this approach through experimental validation on a pre-stressed concrete bridge in Taiwan. It was however observed that lower vehicle speeds provided better results due to higher spectral resolution and reduced interaction with road roughness. 
To enhance the visibility of higher mode bridge frequencies, technique like Empirical Mode Decomposition (EMD)~\cite{huang1998empirical} is used for pre-processing vehicle measurements~\cite{yang2009extraction}. This method aids in extracting more accurate data from the dynamic responses captured using moving sensors. Other studies explored various methods for mode shape identification, such as short-term frequency domain decomposition (STFDD) and Hilbert transform, demonstrating different levels of spatial resolution and effectiveness~\cite{malekjafarian2014identification,yang2014constructing}.
An innovative framework for indirect bridge modal identification, which focuses on extracting higher-mode parameters from vehicle responses through an iterative vehicle response demodulation technique, is detailed in a recent work by Yang et al.~\cite{yang2025indirect}. 
Though, these studies focus on moving sensor problem, it requires multiple sensors and/or maintaining low sensor velocities.
To further investigate the effect of driving frequency induced by the moving vehicles, a detailed study has been conducted by Brewick and Smyth~\cite{brewick2015exploration}.
\par
Researchers have also used the stop-and-go techniques instead of on-the-go techniques to overcome issues associated with the road surface roughness. In the stop-and-go method, the moving sensor/actuator will stop at a location and measure the vibration for a particular time, and then move along the bridge to the next location.
Cara et al. developed a methodology to identify the mode shapes with a roving input and a single sensor using multiple tests~\cite{cara2014estimating,cara2016modal}. Later, Nayek et al.~\cite{nayek2018mass} developed a methodology to identify the mode shapes using a single actuator-sensor pair, including excitation and measurements at different locations along the bridge using either a roving actuator and/or roving sensor. 
In \cite{yang2024bridge}, a dual-axle test vehicle measures the bridge response at designated locations while the vehicle is stationary. The mode shapes are identified using the second-order statistical moment, and the curvature of these mode shapes is utilized to detect damage locations. 
Zhu et al.~\cite{zhu2012wireless} used magnetic wall-climbing robots with the capability of wireless data streaming on a steel truss to identify the mode shapes using the stop-and-go technique. Recently, a study has been done to identify the frequency and mode shape by using programmable wheeled robots in the stop-and-go method~\cite{jian2024robotic}. 
However, in case of continuous traffic flow, the stop-and-go approach may require halting of vehicles to capture the data. Hence, it would be more advantageous to have a method which can identify the bridge modes, with sufficient spatial resolution, on-the-go with a single mobile sensor, while accounting for the randomness induced by traffic loading and road roughness.

In this study, we have proposed two different approaches, for identifying the modal parameters of bridges using vibration data recorded by a single moving sensor. It is illustrated that the modal frequencies and damping ratios can be identified from the acceleration data recorded by the moving sensor using any standard modal identification technique. Then, to identify the mode shapes, two methods are proposed based on the input scenarios.
In the case of known inputs, the identified frequencies and damping ratios are used to obtain the mode shapes, expressed as a linear combination of basic orthonormal polynomials (BOPs), with the coefficients of the BOPs in the linear combinations obtained via optimization. 
In this scenario, the response of the beam is expressed in closed-form using Duhamel's integral.
A statistical approach is then proposed to identify the modal parameters under random traffic loading where exact velocity, arrival time and mass of the vehicles are not known. 
It is shown that the absolute value of the mode shapes are proportional to the ensemble standard deviation (SD) of the modal responses. In this approach, the sensor is required to traverse the bridge multiple times, enabling the identification of mode shapes in both the time and frequency domains, by decomposing measured responses into modal responses. In the time domain, mode shapes are derived from the variances of the modal responses, while in the frequency domain, they are determined by analyzing the evolutionary power spectrum of these responses.
The statistical approach is illustrated using the vehicle-bridge-interaction (VBI) model incorporating the effects of road roughness and randomness in the traffic loading.

The paper is organized as follows. The next section introduces the formulations for the deterministic approach followed by numerical validation considering the bridge modeled as a simply supported beam. In the subsequent section, the formulation for the statistical approach for random traffic loading is developed, followed by numerical illustration. Finally, to incorporate the effect of road roughness, modeled as a homogeneous random field, a vehicle-bridge-interaction (VBI) model is utilized.

\section{Known Input Scenario}{\label{C6sec:method}}
In this section, we develop the formulations for estimating the mode shapes under known input scenario. It is assumed that the frequencies and damping ratios are estimated from the bridge response measured by the moving sensor using any appropriate operational modal analysis technique.
However, since the sensor is moving, it is not possible to identify the mode shapes using any operational modal analysis technique. Hence, alternative formulations are developed to estimate the mode shapes corresponding to the identified frequencies from the response measured using the moving sensor. 


\subsection{Mode Shape Identification}
\subsubsection{Known Input at Particular Location}\label{C6sec:inp_rand}
\begin{figure}[b!]
	\centering
		\includegraphics[scale=.75]{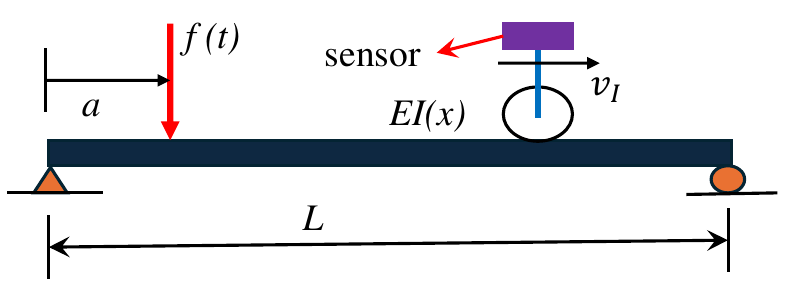}
	   \caption{Simply supported beam with an input force}
	\label{C6FIG:ssb}
\end{figure}

A bridge is modeled as a simply supported Euler-Bernoulli beam of length $L$.
The equation, governing the transverse vibration of the bridge without damping subjected to external dynamic force $f(t)\delta(x-a)$, which is applied at $x=a$ as shown in Fig.~\ref{C6FIG:ssb}, can be written as follows:
\begin{equation}\label{C6eq:EOM}
    m(x) \frac{\partial^2 U(x,t)}{\partial t^2} + \frac{\partial^2}{\partial x^2}\left[ EI(x) \frac{\partial^2 U(x,t)}{\partial x^2} \right] = f(t)\delta(x-a)
\end{equation}
where $m(x)$ and $EI(x)$ are the mass density and flexural rigidity of the bridge, respectively, at position $x$, while $U(x,t)$ denotes the displacement of the bridge. Using modal superposition, the displacement $U(x,t)$ can be written as:
\begin{equation}\label{C6eq:uxt}
    U(x,t) = \sum_{r=1}^{\infty} \phi_r(x) q_r(t)
\end{equation}
where $\phi_r$ and $q_r$ are the mode shape and generalized co-ordinate, respectively, of the $r$th mode.
Incorporating Eq.~\ref{C6eq:uxt} in Eq.~\ref{C6eq:EOM}, and using modal orthogonality, the equation of motion can be written as:
\begin{equation}
    {M}_n \ddot{{q}}_n + {K}_n {q}_n = {F}_n(t); \qquad \textrm{where, } 
    \begin{cases}
        {M}_n = \int_0^L m(x) \left[\phi_n(x) \right]^2 dx; \ \ {F}_n(t) = \int_0^L{f(t) \phi_n(a)}dx \\
        {K}_n = \int_0^L \phi_n(x) \frac{\partial^2}{\partial x^2}\left[ EI(x) \frac{\partial^2 \phi_n(x)}{\partial x^2} \right]  dx
    \end{cases}
\end{equation}
Introducing damping in the equation of motion, it can be re-written as:
\begin{equation}
    {M}_n \ddot{{q}}_n + {C}_n \dot{{q}}_n + {K}_n {q}_n = {F}_n(t); \qquad \textrm{where, } 
        {C}_n = \int_0^L c \left[\phi_n(x) \right]^2 dx = 2\zeta_n \sqrt{M_n K_n}
\end{equation}
where $\zeta_n$ is the damping ratio for the $n$th mode. Now, the equation of motion for any $n$th mode can be written as:
\begin{equation}\label{C6eq:eom_mod}
    \ddot{{q}}_n + 2\zeta_n \omega_n \dot{{q}}_n + \omega_n^2 {q}_n = \frac{f(t)}{{M}_n} \phi_n(a)
\end{equation}
where $\omega_n$ is the natural frequency of the $n$th mode.
A closed-form integral solution of vibration responses in modal co-ordinate can then be obtained from Eq.~\ref{C6eq:eom_mod} as:
\begin{equation}\label{C6eq:ddot_q}
\begin{matrix}
    q_n(t) = \frac{\mathcal{I}_{s_n}}{M_n \omega_{dn}} \phi_n(a); \quad \dot{q}_n(t) = -\zeta_n \omega_n q_n(t) + \frac{\mathcal{I}_{c_n}}{M_n \omega_{dn}} \phi_n(a);\\ \\
    \quad \ddot{q}_n(t) = \frac{f(t)}{{M}_n} \phi_n(a) - 2\zeta_n \omega_n \left[ -\zeta_n \omega_n q_n(t) + \frac{\mathcal{I}_{c_n}}{M_n}\phi_n(a) \right] -\omega_n^2 q_n(t)
    \end{matrix}
\end{equation}
where $\mathcal{I}_{s_n} = \int_0^t{f(\tau) e^{-\zeta_n \omega_n \{t-\tau\} } sin\left[\omega_{dn}\{t-\tau\} \right]d\tau}$ and $\mathcal{I}_{c_n} = \int_0^t{f(\tau) e^{-\zeta_n \omega_n \{t-\tau\} } cos\left[\omega_{dn}\{t-\tau\} \right]d\tau}$; damped natural frequency $\omega_{dn}$ is equal to $\omega_n \sqrt{1-\zeta_n^2}$. 
Using Eqs.~\ref{C6eq:ddot_q} and \ref{C6eq:uxt}, the acceleration response measured by the moving sensor can be calculated as:
\begin{equation}
    \ddot{U}(x_I,t) = \sum_{n=1}^{\infty} \frac{\phi_n(a)}{M_n} \Bigg[ f(t)- \frac{1-2\zeta_n^2}{\sqrt{1-\zeta_n^2}} \omega_n \mathcal{I}_{s_n} - 2\zeta_n \omega_n \mathcal{I}_{c_n} \Bigg] \phi_n(x_I)
\end{equation}
where $x_I$ is the position of the moving sensor. Note that, since the sensor is moving, $x_I$ changes with time as $x_I=v_I t$, where $v_I$ is the velocity of the sensor on the bridge. When $t=0$, the sensor enters the bridge. Thus, $\phi_n(x_I)$ indicates that the mode shapes are evaluated at the position $x_I$ of the moving sensor at time $t$.

\par
The mode shapes of a beam with varying material properties and/or cross-section along the length of the beam can be expressed as a linear combination of orthogonal ploynomials as basis functions. These orthogonal polynomials can be obtained using the Orthonormal Polynomial Series Expansion Method (OPSEM)~\cite{hassanabadi2013new}, based on a Maclaurian series expansion of the displacement of the beam. The Maclaurian series is expressed in terms of Basic Polynomials (BPs), each of which satisfies the boundary conditions of the beam. Then, defining an integral based inner product, an orthonormal set of so-called Basic Orthonormal Polynomials (BOPs) is created by Gram-Schmidt procedure. 
The details of the BOP computation is given in Appendix~\ref{C6append-1}.
The mode shapes in terms of BOPs can be represented as~\cite{hassanabadi2013new}:
\begin{equation}\label{C6eq:ms}
    \phi_n(x_I) = \sum_{i=1}^{N_B} w^n_i \bar{P}_i(x_I)
\end{equation}
where $N_B$ is the number of BOPs considered to represent all the excited mode shapes, and $w_i^n$ is the weight of the $i$th BOP for the $n$th mode. 

To determine the mode shape as a weighted sum of BOPs, we can obtain the weights of these BOPs using any optimization technique,
minimizing the Euclidean norm of the error between the measured acceleration by the moving sensor and a simulated acceleration (with the identified frequencies and damping ratios):
\begin{equation}\label{C6eq:obj}
    \begin{matrix}
        \textrm{arg min } \\ \mathbf{w}
    \end{matrix}   \parallel \ddot{U}(x_I,t) - \hat{\ddot{U}}(\mathbf{w},x_I,t)\parallel_2
\end{equation}
where, $\mathbf{w}=\{w_1^1,\cdots,w_{N_B}^1,w_1^2,\cdots,w_{N_B}^{N_m}\}$ is the weights of the BOPs, and $N_m$ is the number of excited modes. In Eq.~\ref{C6eq:obj}, the simulated $\hat{\ddot{U}}(\mathbf{w},x_I,t)$ can be computed as follows:
\begin{equation}
    \hat{\ddot{U}}(\mathbf{w},x_I,t) = \sum_{n=1}^{\infty} \frac{\sum_{i=1}^{N_B} w^n_i \bar{P}_i(a)}{\int_0^L m(x) \left[\sum_{i=1}^{N_B} w^n_i \bar{P}_i(x) \right]^2 dx} \Bigg[ f(t)- \frac{1-2\zeta_n^2}{\sqrt{1-\zeta_n^2}} \omega_n \mathcal{I}_{s_n} - 2\zeta_n \omega_n \mathcal{I}_{c_n} \Bigg] \sum_{i=1}^{N_B} w^n_i \bar{P}_i(x_I(t))
\end{equation}

\subsubsection{Moving Masses}
In this case, it is considered that the bridge is excited by a moving traffic load, modeled as a series of moving masses. The mass of the vehicles are considered to be negligible compared to the mass of the bridge. Hence, the traffic-bridge system is considered to be a linearly time invariant system.
\begin{figure}[t!]
	\centering
		\includegraphics[scale=.55]{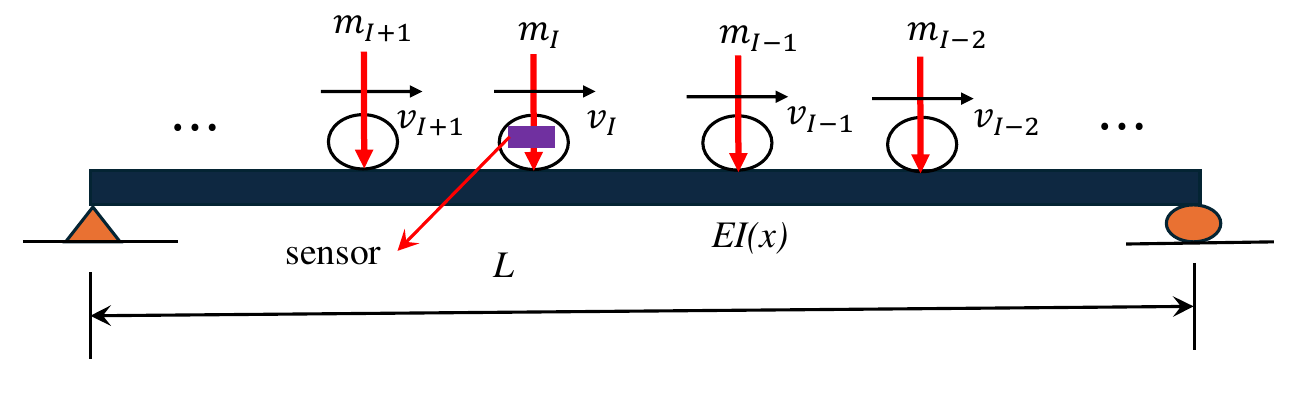}
	   \caption{Simply supported beam with moving masses}
	\label{C6FIG:ssb_mov}
\end{figure}
The position of $i$th moving mass, denoted as $x_i$, will depend on the velocity ($v_i$) and arrival time ($a_i^t$) of that particular moving mass as:
\begin{equation}
    x_i(t) =
    \begin{cases}
        v_i (t-a_i^t) \qquad \textrm{for } 0\leq x_i \leq L \\
        0 \qquad \qquad \quad \textrm{otherwise } 
    \end{cases}
\end{equation}
The acceleration response of the moving sensor can then be written as:
\begin{equation}\label{C6eq:res_mm}
    \ddot{U}(x_I,t) = \sum_{n=1}^{\infty} \sum_{i=1}^{N_v} \frac{\phi_n(x_I)}{M_n} \Bigg[ m_i g \phi_n(x_i)- \frac{1-2\zeta_n^2}{\sqrt{1-\zeta_n^2}} \omega_n \mathcal{I}_{s_n}^{(i)} - 2\zeta_n \omega_n \mathcal{I}_{c_n}^{(i)} \Bigg] \qquad \textrm{where, } 
\end{equation}
where $g$ is the acceleration due to gravity, and 
$\mathcal{I}_{s_n}^{(i)} = \int_0^t{m_i g \{\phi_n \vert _{x_i=v_i\tau}\} e^{-\zeta_n \omega_n \{t-\tau\}} sin\left[\omega_{dn}\{t-\tau\}\right]d\tau}$, and $\mathcal{I}_{c_n}^{(i)} = \int_0^t{m_i g \{\phi_n \vert _{x_i=v_i\tau}\} e^{-\zeta_n \omega_n \{t-\tau\}} cos\left[\omega_{dn}\{t-\tau\}\right]d\tau}$ are the convolution integrals. 
$N_v$ is the total number of masses moving over the bridge, including those currently on/entering the bridge as the sensor traverses it, as well as the masses that have already passed but are still contributing to the free vibration effects.

The mode shapes can be computed using the same approach as given in "Known Input at Particular Location" subsection, 
by solving Eq.~\ref{C6eq:obj}, with $\hat{\ddot{U}}(\mathbf{w},x_I,t)$ in this case expressed as:
\begin{equation}\label{C6eq:u_hat_movMass}
    \hat{\ddot{U}}(\mathbf{w},x_I,t) = \sum_{n=1}^{\infty} \sum_{i=1}^{N_v} \sum_{j=1}^{N_B} \frac{w^n_j \bar{P}_j(x_I)}{\int_0^L m(x) \left[\sum_{j=1}^{N_B} w^n_j \bar{P}_j(x_I) \right]^2 dx} \Bigg[ m_i g \sum_{j=1}^{N_B} w^n_j \bar{P}_j(x_i)- \frac{1-2\zeta_n^2}{\sqrt{1-\zeta_n^2}} \omega_n \mathcal{\widetilde{I}}_{s_n}^{(i)} - 2\zeta_n \omega_n \mathcal{\widetilde{I}}_{c_n}^{(i)} \Bigg] 
\end{equation}
where, the convolution integrals $\mathcal{\widetilde{I}}_{s_n}^{(i)}$ and $\mathcal{\widetilde{I}}_{c_n}^{(i)}$ are represented in terms of BOPs as:
\begin{equation}
\begin{matrix}
    \mathcal{\widetilde{I}}_{s_n}^{(i)} = \sum_{j=1}^{N_B} \left[ \int_0^t{m_i g w^n_j \{ \bar{P}_j \vert _{x_i=v_i\tau}\} e^{-\zeta_n \omega_n \{t-\tau\}} sin\left[\omega_{dn}\{t-\tau\}\right]d\tau} \right]
    \\ \\
    \mathcal{\widetilde{I}}_{c_n}^{(i)} = \sum_{j=1}^{N_B} \left[ \int_0^t{m_i g w^n_j \{\bar{P}_j \vert _{x_i=v_i\tau}\} e^{-\zeta_n \omega_n \{t-\tau\}} cos\left[\omega_{dn}\{t-\tau\}\right]d\tau} \right]
\end{matrix}
\end{equation}



\subsection{Numerical validation}
\subsubsection{Mode Shape Identification Under Random Excitation at Particular Location}
In order to validate the proposed method for known input at a particular location, as given in "KNOWN INPUT SCENARIO" Section, 
we first simulate data based on the following properties of a simply supported beam, serving as a model for the bridge: mass density($m$) = 6.1 Kg/m, cross-section area ($A$)$ = $7.71$\times$ 10$^{-4}$ m$^2$, flexural rigidity ($EI$) $=$ 152.67$\times$ 10$^{3}$ N-m$^2$, and length of the beam ($L$) $=$ 10 m. Using these parameters, the first four natural frequencies of the bridge become 15.61, 62.46, 140.53, and 249.82 rad/s. Rayleigh damping is assumed, with a damping ratio of 2\% for the first four vibration modes.

Gaussian white noise (GWN) is used as input at a point located 2 m (i.e., $L/5$) from the left end of the beam ($a=2$m in Fig.~\ref{C6FIG:ssb}). The weight of the moving sensor is considered to be negligible. This sensor measures the acceleration of the beam while traveling at a velocity of 1 meter per second. The data is sampled at an interval of 0.001 sec, i.e., $dt=0.001$ sec. At any instant of time, the sensor records a single vibration measurement from the beam at a corresponding location. After generating the response of the beam, additive GWN of 5\% root mean squared (rms) of the response is added as measurement noise.

\begin{figure}
	\centering
		\includegraphics[scale=.65]{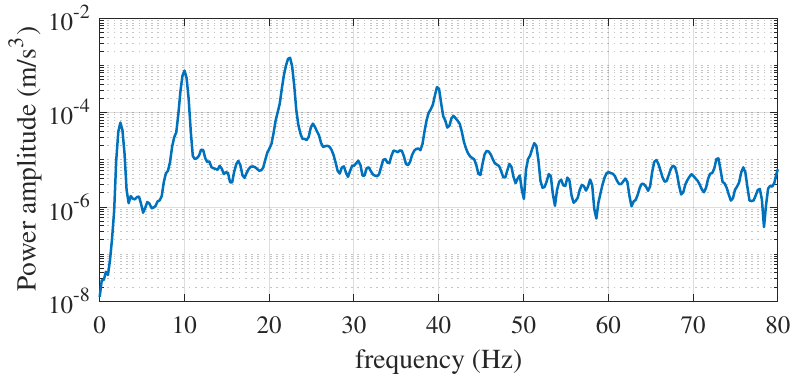}
	   \caption{PSD of the acceleration measured using moving sensor under GWN excitation}
	\label{C6FIG:power_WN}
\end{figure}

\begin{figure}[t!]
	\centering
		\includegraphics[scale=.72]{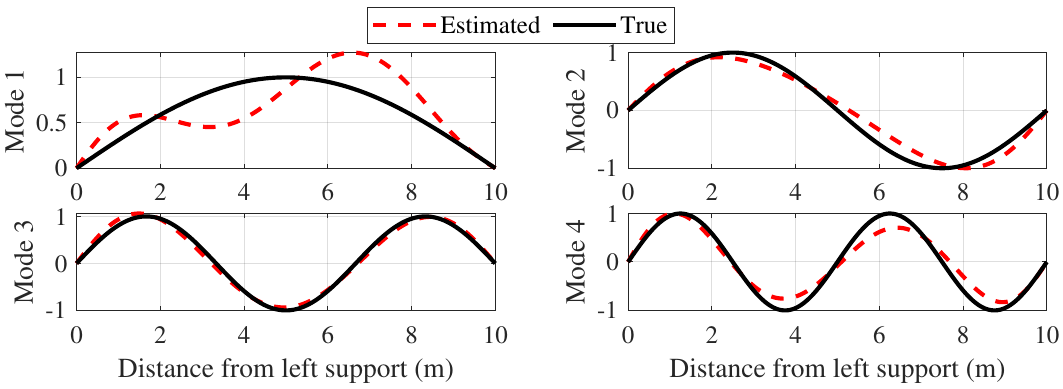}
	   \caption{Identified mode shapes for first four modes of the simply supported under GWN excitation at $L/5$ from left support}
	\label{C6FIG:MS_ID_WN}
\end{figure}

From the power spectrum of the measured acceleration, as shown in Fig.~\ref{C6FIG:power_WN}, it is observed that the first four modes are excited.
The natural frequencies and damping ratios of the beam are identified using Enhanced Frequency Domain Decomposition technique (EFDD)~\cite{brincker2001damping,brincker2001modal} from the measured responses, as detailed in Table~\ref{C6tab:wzeta_WN}.
These identified frequencies and damping ratios are used in $\hat{\ddot{U}}(\mathbf{w},x_I,t)$, which is subsequently used in the objective function specified in Eq.~\ref{C6eq:obj}. Nonlinear least squares, employed through the built-in function $lsqnonlin$ in MATLAB~\cite{MATLAB} with Levenberg-Marquardt algorithm, is used to estimate the weights of BOPs (i.e., $\mathbf{w}$). Once these weights are determined, the mode shapes are computed using Eq.~\ref{C6eq:ms}. The initial weights of BOPs, i.e., $\mathbf{w}$ are set to be zero with the exception of $w_1^1, w_2^2, w_3^3$, and $w_4^4$, which are assigned a value of one.
The estimated mode shapes are presented in Fig.~\ref{C6FIG:MS_ID_WN}. 
The mode shapes of a simply supported beam are given by $sin\left(n\pi x/L \right)$, where $n$ represents the mode number. In this study, these sinusoidal mode shapes of the simply supported beam are considered to be the "true" mode shapes which are also shown in Fig.~\ref{C6FIG:MS_ID_WN} for comparision with the estimated mode shapes.
From these results, it is evident that a single moving sensor is sufficient to identify the mode shapes of the excited modes under a known excitation.
\begin{table}
    \centering
\caption{Identified natural frequency and damping ratio under GWN input }
\label{C6tab:wzeta_WN}
    \begin{tabular}{c|c|c} \hline
  Mode & Natural frequency (rad/s) & Damping ratio (in \%)\\
  \hline
  1 &  15.337&8.94\\
  2 &  62.895&2.60\\
  3&  141.120&1.50\\
  4&  250.008&0.71 \\
     \hline
     \end{tabular}
\end{table}

\begin{figure}
	\centering
		\includegraphics[scale=.65]{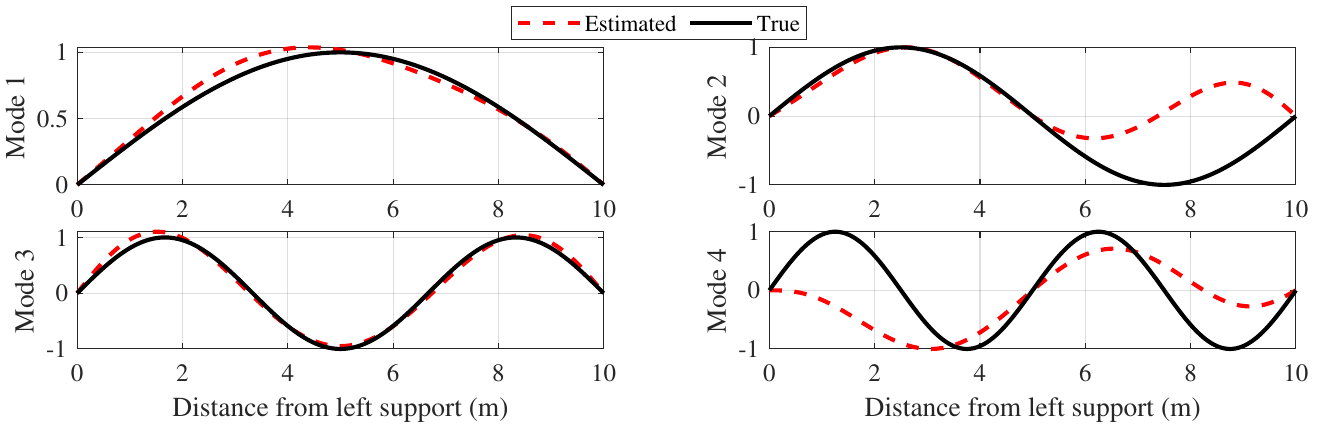}
	   \caption{Identified mode shapes for first four modes of the simply supported under GWN excitation at the centre of the beam}
	\label{C6FIG:MS_ID_WN_Lb2}
\end{figure}

\begin{figure}
	\centering
		\includegraphics[scale=.65]{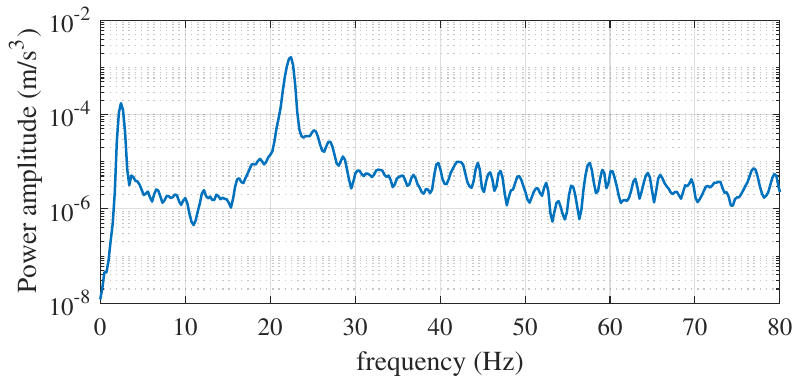}
	   \caption{PSD of the acceleration measured using moving sensor under GWN excitation applied at the centre of the beam}
	\label{C6FIG:power_WN_Lb2}
\end{figure}

It is observed that the estimation of the first mode shape is less accurate compared to the other mode shapes. The disparity in power, as seen in Fig.~\ref{C6FIG:power_WN}, could be the reason in the reduced accuracy in modal identification. The response data collected may not provide sufficient information for accurate identification of the first mode shape, owing to the input location not exciting this mode sufficiently. 
If the same input force is applied at the centre of the beam, the second and fourth mode shapes are not identified, as shown in Fig.~\ref{C6FIG:MS_ID_WN_Lb2} while the first and third mode shapes are estimated accurately. In this case, it can be observed from the PSD plot of the response of the moving sensor in Fig.~\ref{C6FIG:power_WN_Lb2} that, due to the input position, the second and fourth modes are not excited. This has also been shown in a previous study~\cite{jana2019fisher} that input force locations can significantly enhance the quality of measurement data for identified/excited modes for some input locations or unidentified modes for some specific input locations.


\subsubsection{Mode Shape Identification Under Moving Masses}
In this numerical illustration, we analyze the same bridge model, as discussed in the previous subsection, subjected to vibrations caused by moving masses. We consider a total of five moving masses ($N_v=5$), each weighing 1 kg. This mass is significantly smaller than the overall mass of the beam (i.e., 61 kg), thereby the system is assumed to be time-invariant.
The velocities of the moving masses are specified as follows: 1 m/s, 0.5 m/s, 2 m/s, 1.5 m/s, and 4 m/s, respectively. The second mass is equipped with an accelerometer to measure the vibrations of the bridge; hence, its velocity is intentionally reduced compared to the others to enhance the spatial density of the measurements.
Additionally, a time lag of 1 second is considered between the arrival of each consecutive mass onto the bridge. 
The dynamic response of the bridge to the vibrations induced by the moving masses, and measured by the moving sensor is then simulated.
\begin{table}
    \centering
\caption{Identified natural frequency and damping ratio in moving mass scenario }
\label{C6tab:wzeta_mmas}
    \begin{tabular}{c|c|c} \hline
  Mode & Natural frequency (rad/s) & Damping ratio (in \%)\\
  \hline
  1 &  15.337& 4.97\\
  2 &  62.329& 1.69\\
  3& 138.104& 0.49\\
  4& 251.202& 0.27\\
     \hline
     \end{tabular}
\end{table}

\begin{figure}
	\centering
		\includegraphics[scale=.65]{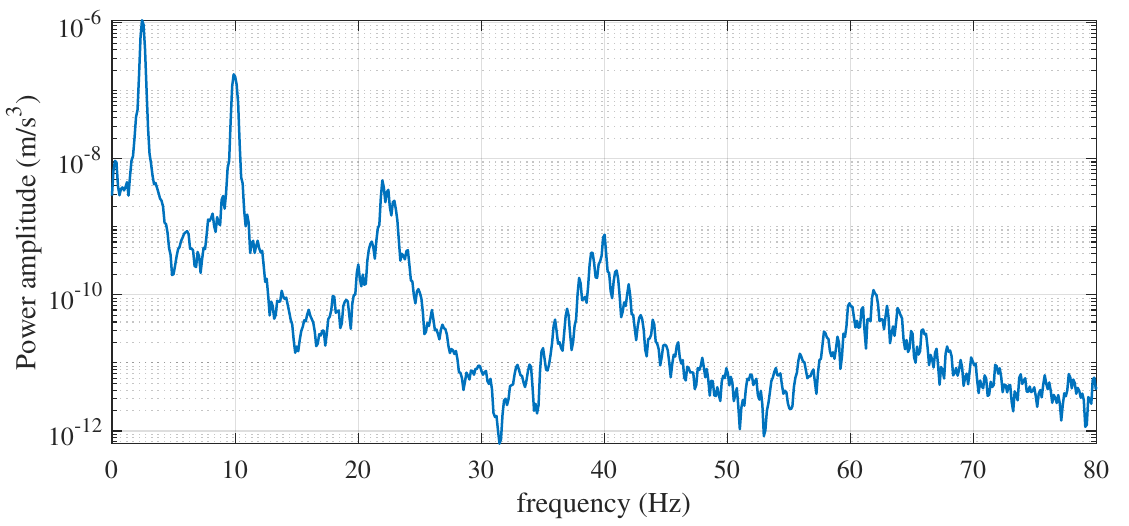}
	   \caption{PSD of the acceleration measured using moving sensor in moving mass scenario}
	\label{C6FIG:power_mmas}
\end{figure}

\begin{figure}
	\centering
		\includegraphics[scale=.65]{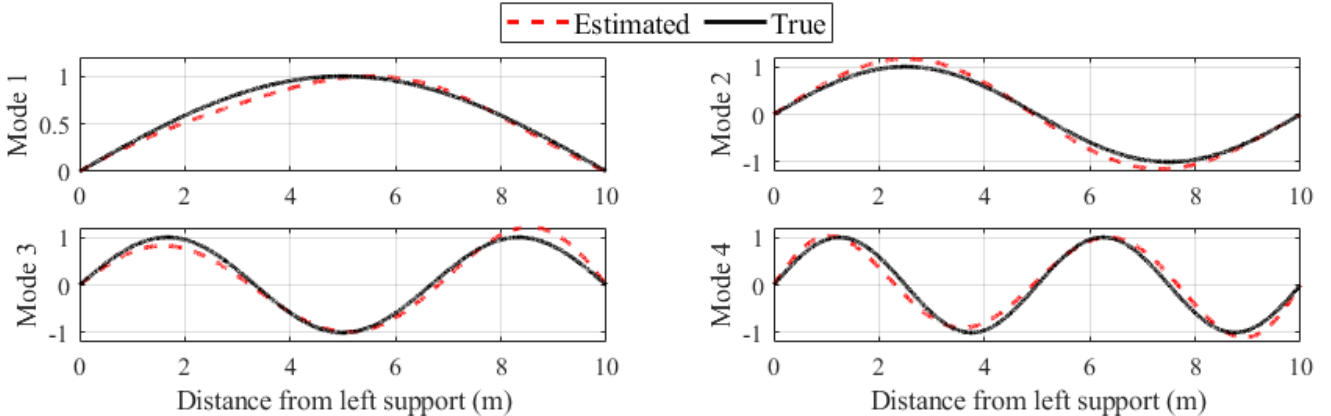}
	   \caption{Identified mode shapes for first four modes of the simply supported beam in moving mass scenario}
	\label{C6FIG:MS_ID_mmas}
\end{figure}

 The power spectrum of the measured acceleration of the bridge, as illustrated in Fig.~\ref{C6FIG:power_mmas}, shows substantial power in the first four modes indicating successful excitation of the first four modes under traffic loading (moving mass input). The natural frequencies and damping ratios are estimated using EFDD method, and these identified values are given in Table~\ref{C6tab:wzeta_mmas} for the first four modes. The weights of the BOPs are estimated by solving the optimization problem, as given in Eq.~\ref{C6eq:obj}, using the same technique as discussed in the previous subsection. These weights are subsequently applied in Eq.\ref{C6eq:ms} to compute the mode shapes. From Fig.~\ref{C6FIG:MS_ID_mmas}, it is evident that all the four modes can be successfully identified. Compared with the results of the previous sub-section, the findings suggest that mode shape identification under traffic loading is more effective than exciting with an actuator at a specific location on the bridge owing to some modes not being sufficiently excited for different fixed input locations.



\section{Statistical Approach} \label{C6sec:stat_formu}
In the case of continuous traffic flow, it may be difficult to ascertain the mass, velocity and arrival times of the vehicles. While the velocity and arrival times may be obtained via video measurements, precise information about the mass of vehicles will not be known in general. To tackle this challenge, a statistical approach is developed in this section to effectively identify the mode shapes corresponding to the excited modes of the bridge in case of unknown random traffic loading. The approach is developed both in time as well as frequency domains.
This approach requires the moving sensor to traverse the bridge multiple times, in the same speed to ensure temporal consistency of the measured data in each run of the moving sensor.
\subsection{Formulation}
\subsubsection{Mode shape estimation using variances} \label{C6sec:stat_appro}
Using modal superposition, the response measured by the moving sensor, $\ddot{U}(x_I,t)$ can be represented as the sum of the modal responses as follows:
\begin{equation}
    \ddot{U}(x_I,t) = \ddot{u}_1(x_I,t) + \ddot{u}_2(x_I,t) + \ddot{u}_3(x_I,t) + \cdots 
\end{equation}
where decomposed response for the $i$th mode $\ddot{u}_i(x_I,t)$, can be written using Eq.~\ref{C6eq:res_mm}, as:
\begin{equation}\label{C6eq:acc_IV_mode}
    \ddot{u}_n(x_I,t) = \phi_n(x_I) \ddot{q}_n (t); \quad \textrm{where, }
    \ddot{q}_n (t) = \sum_{i=1}^{N_v} \frac{1}{M_n} \left( m_i \phi_n(x_i)- \frac{1-2\zeta_n^2}{\sqrt{1-\zeta_n^2}} \omega_n \mathcal{I}_{s_n}^{(i)} - 2\zeta_n \omega_n \mathcal{I}_{c_n}^{(i)} \right) 
\end{equation}
It is assumed that the modal responses can be obtained from the measured (total) response by sensor using an appropriate modal decomposition technique (such as empirical mode decomposition (EMD) technique~\cite{huang1998empirical}.

Theoretically the variance of $\ddot{u}_n$ at each position along the bridge can be expressed as:
\begin{equation}
    \sigma_{\ddot{u}_n}^2 (x_I,t) = \mathsf{E} \left[ \left( \ddot{u}_n(x_I,t) - \mathsf{E}\left[ \ddot{u}_n(x_I,t) \right] \right)^2 \right] 
\end{equation}
where $\mathsf{E}$ is the expectation operator. As the position of the moving sensor is known, using Eq.~\ref{C6eq:acc_IV_mode}, $\sigma_{\ddot{u}_n}^2 (x_I,t)$ can be written as:
\begin{equation}\label{C6eq:var_1}
    \sigma_{\ddot{u}_n}^2 (x_I,t) = \phi^2_n(x_I)  \mathsf{E} \left[ \left( \ddot{q}_n(t) - \mathsf{E}\left[ \ddot{q}_n(t) \right] \right)^2 \right] = \phi^2_n(x_I) \sigma^2_{\ddot{q}_n} (t)
\end{equation}
The arrival time of the traffic flow typically follows a Poisson distribution~\cite{hosamo2012study,gerlough1955use}. In case of stationarity of the occurrence rate of the Poisson distribution, $\sigma_{{q}_n} (t)$ becomes independent of time as demonstrated in~\cite{sniady1984vibration,iwankiewicz1984vibration}. Additionally, in Appendix~\ref{C6append-2}, it is shown that $\sigma_{\ddot{q}_n} (t)$ is also time-independent under the condition of a stationary arrival rate of the moving masses.
Therefore, Eq.~\ref{C6eq:var_1} can be written as:
\begin{equation}\label{C6eq:sig_MS}
    \sigma_{\ddot{u}_n}^2 (x_I,t) = \phi^2_n(x_I) {\sigma}^2_{\ddot{q}_n}; \qquad \textrm{or,} \qquad  \sigma_{\ddot{u}_n}^2 (x_I,t) \propto \phi^2_n(x_I)
\end{equation}
Thus, the square of the mode shapes can be estimated for the excited modes, from which the mode shapes may then be obtained using appropriate sign.

\subsubsection{Mode shape estimation using evolutionary power spectrum}\label{C6sec:stat_appro_EPS}
The pattern of the mode shapes can also be determined in the frequency domain by computing the evolutionary power spectrum (EPS) of the modal responses. First, autocorrelation matrix ($\mathbf{R}_{\ddot{u}_i}$) for the $i$th modal response is computed as follows:
\begin{equation}\label{C6eq:corr_eps}
    \mathbf{R}_{\ddot{u}_i} = 
    \begin{pmatrix}
        \mathsf{E}[\ddot{u}_i(\Delta t),\ddot{u}_i(\Delta t)], & \mathsf{E}[\ddot{u}_i(\Delta t),\ddot{u}_i(2\Delta t)],  & \cdots & \mathsf{E}[\ddot{u}_i(\Delta t),\ddot{u}_i(N_T\Delta t)] \\ 
        \mathsf{E}[\ddot{u}_i(2\Delta t),\ddot{u}_i(\Delta t)], & \mathsf{E}[\ddot{u}_i(2\Delta t),\ddot{u}_i(2\Delta t)], & \cdots & \mathsf{E}[\ddot{u}_i(2\Delta t),\ddot{u}_i(N_T\Delta t)]\\ 
        \vdots & \vdots & \vdots & \ddots & \vdots \\ 
        \mathsf{E}[\ddot{u}_i(N_T\Delta t),\ddot{u}_i(\Delta t)], & \mathsf{E}[\ddot{u}_i(N_T\Delta t),\ddot{u}_i(2\Delta t)], & \cdots & \mathsf{E}[\ddot{u}_i(N_T\Delta t),\ddot{u}_i(N_T\Delta t)]
    \end{pmatrix}
\end{equation}
where $N_T$ denotes the total number of data points collected by the sensor when traversing the bridge. Each column of this matrix represents the correlation of measurements taken at a specific location with all other measured locations on the bridge, indicating how the measurements are spatially correlated across the structure. By performing the Fourier transform for each column, the power spectrum may be computed as follows:
\begin{equation}\label{C6eq:powerSpec_eps}
    \mathbf{S}_{\ddot{u}_i}(\omega,t) = 
    \begin{pmatrix}
        S_{\ddot{u}_i}(\Delta\omega,\Delta t), & S_{\ddot{u}_i}(\Delta\omega,2\Delta t), & \cdots & S_{\ddot{u}_i}(\Delta\omega,N_T \Delta t) \\
        S_{\ddot{u}_i}(2\Delta\omega,\Delta t), & S_{\ddot{u}_i}(2\Delta\omega,2\Delta t), & \cdots & S_{\ddot{u}_i}(2\Delta\omega,N_T \Delta t) \\
        \vdots & \vdots & \vdots & \ddots & \vdots \\
        S_{\ddot{u}_i}(N_\omega \Delta\omega,\Delta t), & S_{\ddot{u}_i}(N_\omega \Delta\omega,2\Delta t), & \cdots & S_{\ddot{u}_i}(N_\omega \Delta\omega,N_T \Delta t)
    \end{pmatrix}
\end{equation}
where $N_\omega$ denotes length of the vector containing the frequencies after the Fourier transform. Thus, $\mathbf{S}_{\ddot{u}_i}$ becomes an evolutionary power spectrum (EPS) due to the dependence on both time and frequency, which captures how power distribution evolves across frequencies over time. 

To compute the power spectrum for each column, half of the total data points, $N_h$ (i.e., $N_T/2$ (for even row numbers) or $(N_T-1)/2$ (for odd row numbers)), are selected. The vector for each column is defined as follows:
\begin{equation}
    \mathbf{R}_v^j = 
    \begin{cases}
        \mathbf{R}_{\ddot{u}_i}([j:j+N_h-1],j) \qquad  \textrm{for }j\leq N_h \\
        \mathbf{R}_{\ddot{u}_i}([j-N_h+1:j],j) \qquad \textrm{for }j> N_h
    \end{cases}
\end{equation}
where $\mathbf{R}_{\ddot{u}_i}(.)$ denotes the rows selected for the $j$th column. Selecting only the half number of data points of the autocorrelation ($\mathbf{R}_{\ddot{u}_i}$) does not change the properties as it is an even function. 
Given that $\ddot{u}_i(x_I,t)=\phi_i(x_I) \ddot{q}_i(t)$), and noting the stationarity of $\ddot{q}_i(t)$, power $S_{\ddot{u}_i}(\omega,t)$ for a particular frequency (say, $\omega=\omega_k$) becomes directly proportional to the square of the $i$th mode shape as follows:
\begin{equation}
    S_{\ddot{u}_i}(\omega_k,t) \propto \phi_i^2(x_I(t))
\end{equation}
Thus, the square of mode shape for $i$th mode can be obtained as the particular row, corresponding to the frequency of the $i$th mode, of the power spectrum matrix. 

\subsection{Numerical Simulation with moving masses}\label{C6sec:stat_num_mmas}
To illustrate the approach, the same bridge model as used in the previous sections is considered. The analysis takes into account 50 separate runs of the moving sensor along the bridge.
A total of 25  moving masses are considered, with a mean velocity of 2 m/s and a mean mass of 1 kg. 
The moving sensor, with a mass of 1 kg, travels at a constant velocity of 0.5 m/s for all the 50 simulations.
The position of the moving sensor is chosen in such a way that during the time the sensor is on the bridge, a continuous flow of traffic is maintained.
The masses and velocities of the moving masses are assumed to follow uniform distributions in the intervals of $\pm$20\% and $\pm$2.5\% around their respective means. Additionally, the arrival times of the moving masses are modeled using a Poisson distribution with an occurrence rate of 1 to capture the randomness of the vehicle arrivals~\cite{hosamo2012study,gerlough1955use}.

\begin{figure}
	\centering
		\includegraphics[scale=.65]{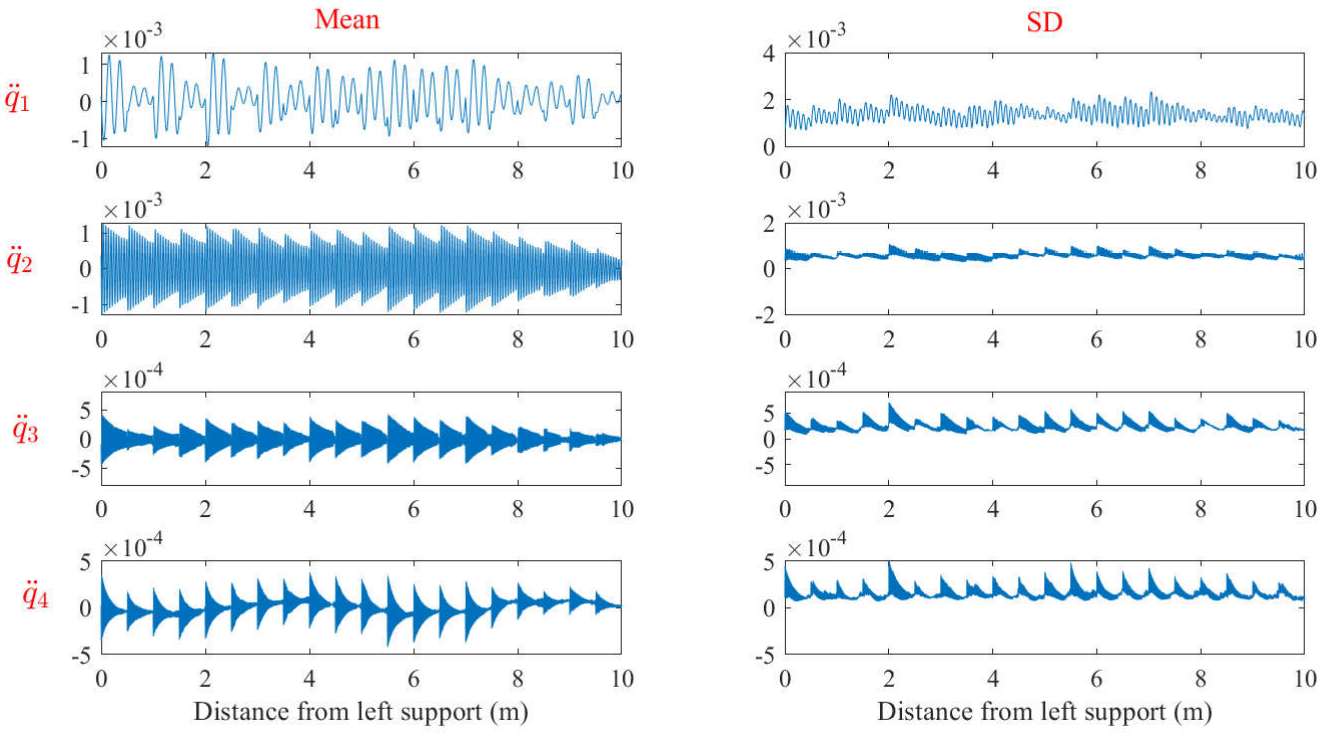}
	   \caption{Variation in mean and standard deviation (SD) of the simulated responses $\ddot{q}_1$ to $\ddot{q}_4$ along the length of the bridge}
	\label{C6FIG:mmas_stat_muSig}
\end{figure}

\textbf{Validation using variances: }
From the simulated $\ddot{q}_i$, it is observed that while the mean values may vary, the standard deviation (SD) of $\ddot{q}_i$ remains relatively constant at each point along the bridge for a given type of traffic loading, as illustrated in Fig.~\ref{C6FIG:mmas_stat_muSig}. 
Since the SD of the $\ddot{q}_i$ is nearly constant, the variances of the modal responses (i.e., $\ddot{u}_i$) is proportional to the square of the mode shapes at each measurement location, as indicated in Eq.~\ref{C6eq:sig_MS}. 

\begin{figure}
	\centering
		\includegraphics[scale=.75]{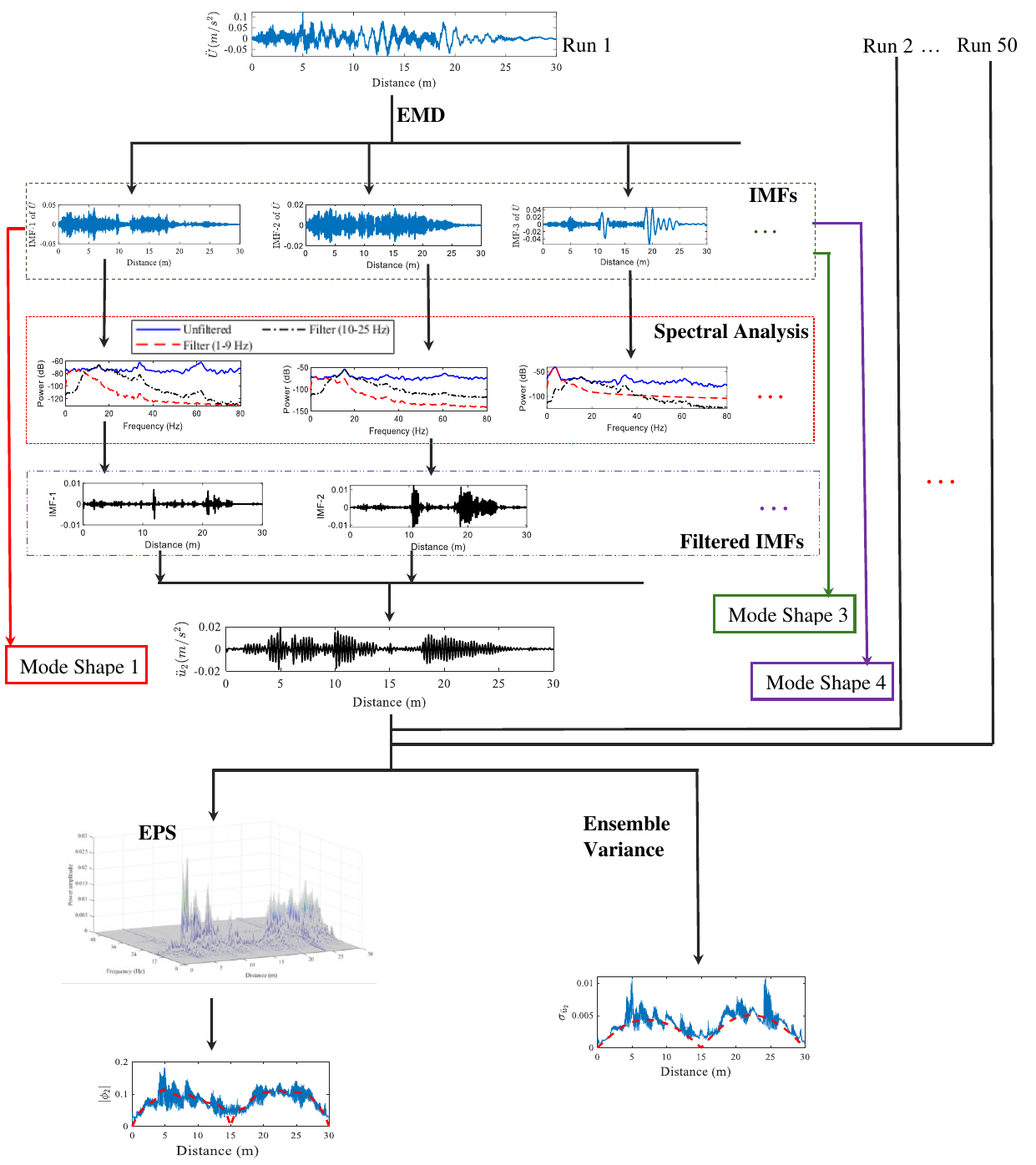}
	   \caption{Flowchart for mode shape identification by using the decomposed modal responses}
	\label{C6FIG:MS_id_flowchart_EMD}
\end{figure}

\begin{figure}
	\centering
		\includegraphics[scale=.65]{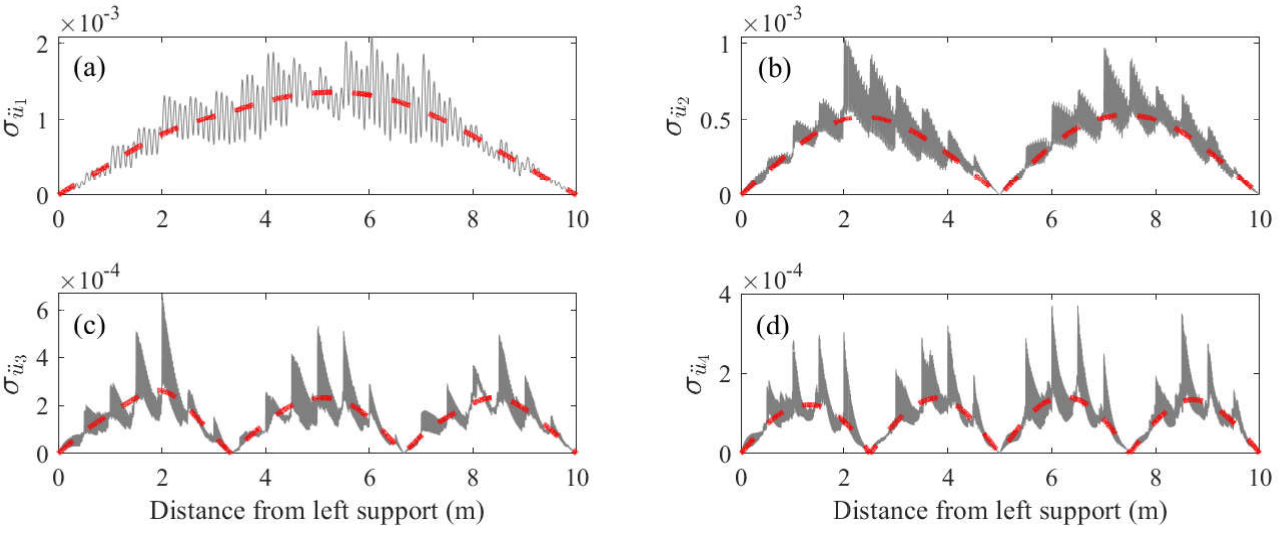}
	   \caption{SD of the simulated modal responses $\ddot{u}_1$ to $\ddot{u}_4$ in (a-d), respectively, along the length of the bridge}
	\label{C6FIG:mmas_stat_mshape}
\end{figure}

The frequencies and damping ratios are identified using EFDD method, and these values are almost the same (maximum error within $\pm 0.10\%$ in frequencies and $\pm 2.50\%$ in damping ratios) as shown in Table~\ref{C6tab:wzeta_mmas}. To identify the mode shapes corresponding to these frequencies using the response from a single moving sensor, without knowing the exact mass and velocity of the vehicles, we have used statistical approach as explained in "Mode shape estimation using variances" subsection. 
To compute the mode shape patterns as given in Eq.~\ref{C6eq:sig_MS}, the measured responses must be decomposed into their modal components. For this decomposition, we have utilized the empirical mode decomposition (EMD) technique~\cite{huang1998empirical}, which decomposes the original response into intrinsic mode functions (IMFs). By performing spectral analysis on these estimated IMFs, the modal reponses are computed by summing those IMFs that contain the same frequency content. In cases where certain IMFs contain multiple frequency components, a bandpass filter is applied to further isolate these components.
These filtered IMFs are then used to obtain the modal response of the corresponding mode, as shown in the flowchart in Fig.~\ref{C6FIG:MS_id_flowchart_EMD}~\footnote{In this flowchart, the responses from the TR-I example of the vehicle-bridge-interaction model are utilized for illustration purposes}. For example, the spectral analysis in Fig.~\ref{C6FIG:MS_id_flowchart_EMD} shows that the third IMF does not have any components of second mode; therefore it is not considered in the estimation of second modal response, i.e., $\ddot{u}_2$.
Following the decomposition of the responses for each run of the vehicle, the ensemble standard deviation (SD) of those decomposed responses are
computed. The SD of the modal responses ($\sigma_{\ddot{u}_i}$) corresponding to different frequencies are shown in Fig.~\ref{C6FIG:mmas_stat_mshape}. To enhance the clarity of the identified mode shapes, a three point moving average filter, with 25\% weightage for 1st and 3rd point while 50\% weightage for the second (middle) point, is applied to smooth the results (shown with thick dotted lines in Fig.~\ref{C6FIG:mmas_stat_mshape}), providing a clearer representation of the mode shape patterns. 
From such an identification, by leveraging engineering knowledge, the sign of the mode shapes corresponding to the identified frequencies can be estimated. 

\begin{figure}
	\centering
		\includegraphics[scale=.65]{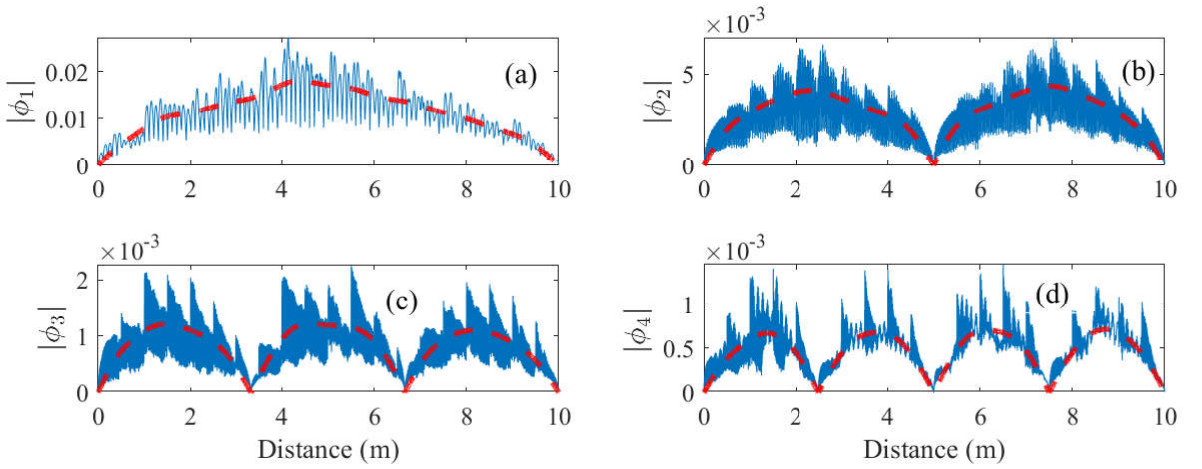}
	   \caption{Absolute value of mode shapes for the first four modes in (a-d), respectively, along the length of the bridge}
	\label{C6FIG:MS_EPS_mmas}
\end{figure}

\textbf{Validation using EPS: }In this case, the magnitude of the mode shapes is determined through the evolutionary power spectrum (EPS) approach, as mentioned in "Mode shape estimation using evolutionary power spectrum" subsection. 
Firstly, EPS is calculated for each modal response, which is derived using the EMD technique as explained previously. The row of the EPS provides the squared values of the mode shapes for the respective mode. The magnitude of the mode shapes is then obtained by taking the square root of these values, which are shown in Fig.~\ref{C6FIG:MS_EPS_mmas}, where thick dotted lines denote the moving average values.

\section{Numerical validation for Statistical Approach with VBI model}\label{C6sec:stat_appro_valid}
In this section, a more complex scenario involving the effect of road roughness is explored by employing a VBI model. A spring-damper-mass model has been considered to simulate the vehicle dynamics effectively. 
When a sensor is installed on the vehicle, the bridge responses can be extracted from the vehicle response knowing the vehicle dynamics, e.g., the vehicle transfer function, and obtaining the bridge response (as in an input to the vehicle) from the vehicle response (output) using the known vehicle transfer function/dynamics properties. 
However, in this paper, we assume that the sensor is positioned on the vehicle axle and ignore any wheel deformation.
These assumptions imply that the measured acceleration by the sensor is equal to the response of the bridge.
The following assumptions are made in this VBI model:
\begin{enumerate}
    \item sensor is installed on the axle of the vehicle
    \item deformation of tires and wheels are ignored for simplicity
    \item tires are always in contact with the road surface
\end{enumerate}

\begin{figure}
	\centering
		\includegraphics[scale=.65]{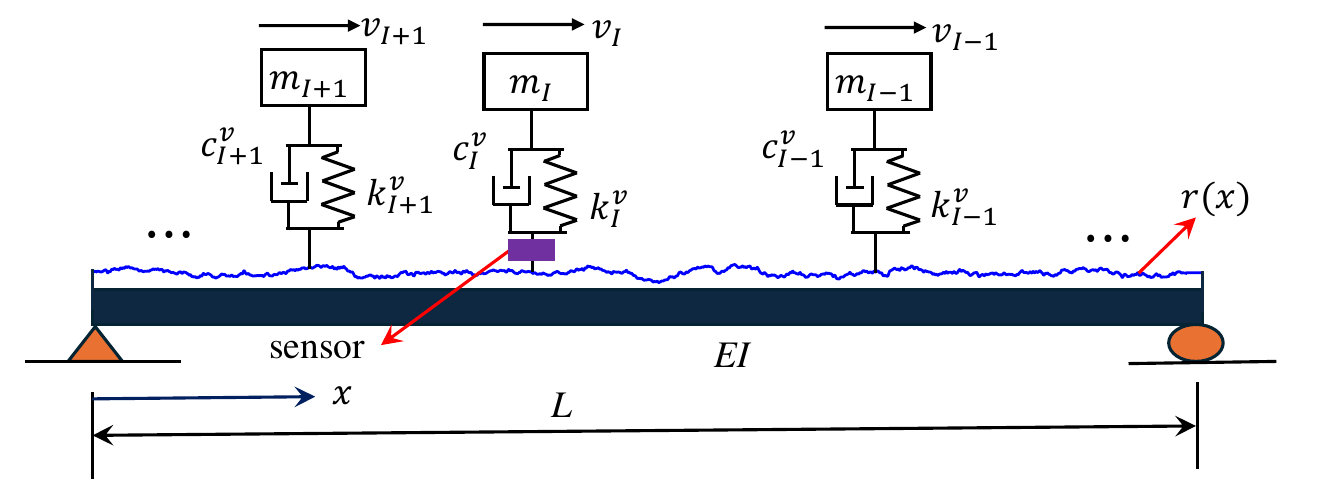}
	   \caption{Simply supported beam with moving spring-damper-mass model of vehicles with road roughness}
	\label{C6FIG:vbi_rx}
\end{figure}
The equation of $j$th spring-damper-mass model (representing the vehicle) on the $e$th beam element in the VBI model, is given by~\cite{yang2018damping}:
\begin{equation}\label{C6eq:vbi_1}
    m_j^v \ddot{y}_j^v + c_j^v \frac{d}{dt} \left( y_j^v -\left(\mathbf{N}_e^c\right)^\top \mathbf{y}_e^b -r_c \right) + k_j^v \left( y_j^v -\left(\mathbf{N}_e^c\right)^\top \mathbf{y}_e^b -r_c \right) = 0 
\end{equation}
where $m_j^v$, $c_j^v$ and $k_j^v$ denote mass, damping and stiffness of the $j$th spring-damper-mass (vehicle) model; $y_j^v$, $\dot{y}_j^v$, and $\ddot{y}_j^v$ are the displacement, velocity and acceleration of the $j$th vehicle.
The road roughness of the bridge span is denoted as $r(x)$, with $r_c=r(x=x_c)$, indicating the roughness at the contact point $x_c$.
The equation for the beam element is:
\begin{equation}\label{C6eq:vbi_2}
    \mathbf{m}_e^b \ddot{\mathbf{y}}_e^b + \mathbf{c}_e^b \dot{\mathbf{y}}_e^b + \mathbf{k}_e^b \mathbf{y}_e^b = f_j^c \mathbf{N}_e^c
\end{equation}
where $\mathbf{y}_e^b$, $\dot{\mathbf{y}}_e^b$, and $\ddot{\mathbf{y}}_e^b$ are the displacement, velocity and acceleration, of the beam element, respectively.
The matrices $\mathbf{m}_e^b$, $\mathbf{k}_e^b$ and $\mathbf{c}_e^b$ correspond to the mass, stiffness and damping of the beam element. 
The shape function for the beam element, $\mathbf{N}_e(x)$ is expressed using Hermite polynomials as:
\begin{equation}
    \mathbf{N}_e(x) = \left[ 1-3\frac{x^2}{L^2}+ 2\frac{x^3}{L^3},\,
        x-2\frac{x^2}{L},\, 3\frac{x^2}{L^2}-2\frac{x^3}{L^3}, \,
        \frac{x^3}{L^2}-\frac{x^2}{L} \right]^\top
\end{equation}

The contact force between the bridge and vehicle, $f_j^c$, can be computed as:
\begin{equation}\label{C6eq:fcj}
    f_j^c = -m_j^v g + c_j^v \frac{d}{dt} \left( y^v -\left(\mathbf{N}_e^c\right)^\top y^b -r_c \right) + k_j^v \left( y^v -\left(\mathbf{N}_e^c\right)^\top y^b -r_c \right)
\end{equation}
In Eq.~\ref{C6eq:fcj}, $\mathbf{N}_e^c$ is the vector containing the values of the shape function at the contact point, i.e., $\mathbf{N}_e(x)$ evaluated at $x=x_c$.

The system of equations for a VBI element can be derived by combining Eq.~\ref{C6eq:vbi_1} and \ref{C6eq:vbi_2} as follows:
\begin{equation}
\begin{matrix}
    \begin{bmatrix}
        m_j^v & 0 \\ 0 & \mathbf{m}_e^j
    \end{bmatrix}
    \begin{Bmatrix}
        \ddot{y}_j^v \\ \ddot{\mathbf{y}}_e^b
    \end{Bmatrix}
    + 
    \begin{bmatrix}
        c_j^v & -c_j^v \left(\mathbf{N}_e^c\right)^\top \\ -c_j^v \mathbf{N}_e^c & \mathbf{c}_e^b + c_j^v \left(\mathbf{N}_e^c\right) \left(\mathbf{N}_e^c\right)^\top 
    \end{bmatrix} 
    \begin{Bmatrix}
        \dot{y}_j^v \\ \dot{\mathbf{y}}_e^b
    \end{Bmatrix}
    +
    \\
    \begin{bmatrix}
        k_j^v & -c_j^v v_j {\left(\mathbf{N}_e^c\right)^\prime}^\top - k_j^v \left(\mathbf{N}_e^c\right)^\top \\ -k_j^v \mathbf{N}_e^c & \mathbf{k}_e^b + c_j^v v_j \left(\mathbf{N}_e^c\right) {\left(\mathbf{N}_e^c\right)^\prime}^\top + k_j^v \left(\mathbf{N}_e^c\right) \left(\mathbf{N}_e^c\right)^\top 
    \end{bmatrix}
    \begin{Bmatrix}
        {y}_j^v \\ {\mathbf{y}}_e^b
    \end{Bmatrix}
    = 
    \begin{Bmatrix}
        c_j^v v_j r^\prime_c + k_j^v r_c \\ 
        - c_j^v v_j r^\prime_c \left(\mathbf{N}_e^c\right) - k_j^v r_c \mathbf{N}_e^c - m_j^v g \mathbf{N}_e^c
    \end{Bmatrix}
\end{matrix}
\end{equation}
where (.)$^\prime$ indicates the differentiation with respect to $x$, while the superscript $\top$ represents the transpose of a vector/matrix.
To extend this formulation for multiple vehicles passing over the bridge, the equation of motion can be written as:
\begin{equation}
\begin{matrix}
    \begin{bmatrix}
        \mathbf{M}^v & 0 \\ 0 & \mathbf{M^b}
    \end{bmatrix}
    \begin{Bmatrix}
        \ddot{\mathbf{y}}^v \\ \ddot{\mathbf{y}}^b
    \end{Bmatrix}
    + 
    \begin{bmatrix}
        \mathbf{C}^v & -\mathbf{C}^v \mathbf{N}_G^\top \\ -\mathbf{C}^v \mathbf{N}_G & \mathbf{C}^b + \mathbf{N}_G \mathbf{C}^v \mathbf{N}_G^\top 
    \end{bmatrix} 
    \begin{Bmatrix}
        \dot{\mathbf{y}}^v \\ \dot{\mathbf{y}}^b
    \end{Bmatrix}
    +
    \begin{bmatrix}
        \mathbf{K}^v & -\mathbf{C}^v \mathbf{v} {\mathbf{N}_G^\prime}^\top - \mathbf{K}^v \mathbf{N}_G^\top \\ -\mathbf{K}^v \mathbf{N}_G & \mathbf{K}^b + \mathbf{N}_G \mathbf{C}^v \mathbf{v} {\mathbf{N}_G^\prime}^\top + \mathbf{N}_G \mathbf{K}^v \mathbf{N}_G^\top 
    \end{bmatrix}
    \begin{Bmatrix}
        {\mathbf{y}}^v \\ {\mathbf{y}}^b
    \end{Bmatrix}
    \\
    =
    \begin{Bmatrix}
        \mathbf{C}^v \mathbf{v} \mathbf{r}^\prime_G + \mathbf{K}^v \mathbf{r}_G \\ 
        - \mathbf{N}_G \mathbf{C}^v \mathbf{v} \mathbf{r}^\prime_G - \mathbf{N}_G \mathbf{K}^v \mathbf{r}_G - \mathbf{N}_G  \mathbf{v}  \mathbf{g}
    \end{Bmatrix}
\end{matrix}
\end{equation}
where $\mathbf{M^b}$, $\mathbf{K^b}$, and $\mathbf{C^b}$ represent the global mass, stiffness and damping matrices of the beam, respectively,  with a size of ($N\times N$). The displacement, velocity and acceleration responses of all active degrees of freedom (DOFs) (i.e., $N$) of the FE model of the beam are $\mathbf{y}^b$, $\dot{\mathbf{y}}^b$, and $\ddot{\mathbf{y}}^b$, respectively. $\mathbf{v}$ is a diagonal matrix of size $N_v\times N_v$ consisting of the velocities of all the vehicles in the diagonal elements. Similarly, $\mathbf{M}^v$, $\mathbf{K}^v$, and $\mathbf{C}^v$ are diagonal matrices of size $N_v\times N_v$ containing the masses, stiffnesses and damping coefficients, respectively, of the spring-damper-mass models. The acceleration due to gravity is denoted as $g$, and $\mathbf{g}$ is a vector of size $N_v\times 1$ with all the values equal to $g$.
The matrix, $\mathbf{N}_G$ consists the shape function values at the contact points of all the vehicles at any particular time $t$, expresses as:
\begin{equation}
    \mathbf{N}_G = 
    \begin{bmatrix}
        \mathbf{N}_c^1(t) & \mathbf{N}_c^2(t) & \cdots & \mathbf{N}_c^{N_v}(t) 
    \end{bmatrix}_{N \times N_v}
\end{equation}
where $\mathbf{N}_c^j(t)$ contains the values of the shape function corresponding to the contact point of the $j$th vehicle to the bridge at time $t$. The vector $\mathbf{r}_G$, of size $N_v\times 1$, contains the corresponding values of road roughness at the contact points of the vehicles. The details of the road roughness profile is discussed next.

\par
\textbf{Road Roughness:}
The road roughness is used here to represent the roughness of the surface of the bridge. To numerically simulate the road roughness $r(x)$, we model the roughness as in~\cite{yang2018damping}:
\begin{equation}
    r(x) = \sum_i d_i cos(\kappa_i x_c + \theta_i)
\end{equation}
 where $\kappa_i$ denotes the $i$th spatial frequency, ranging from 1 to 100 cycle/m with an increment $\Delta \kappa=$ 0.04 cycle/m; $\theta_i$ represents the random phase angle. The amplitude $d_i$ for each roughness class can be computed using:
 \begin{equation}
     d_i = \sqrt{2G_d(\kappa_i)\Delta \kappa}
 \end{equation}
 where the power spectral density (PSD) function is defined as\cite{technical1995mechanical}:
 \begin{equation}
     G_d(\kappa_i) = G_d(\kappa_0) \left( \frac{\kappa_i}{\kappa_0} \right)^{-w}
 \end{equation}
with $w=$ 2 and $\kappa_0=$ 0.1 cycle/m; and $G_d(\kappa_0)$ is calculated based on the roughness class.

\par
\textbf{Numerical Validation:}
To validate the formulation, a numerical simulation is conducted. A bridge of 30 m span is considered, with a modulus of elasticity, $E$ = 27.5 GPa, mass density = 1000 kg/m, and moment of inertia of the cross section = 0.175 m$^4$. 
Rayleigh damping is assumed with 1\% damping ratio for the first 6 modes of the bridge. First four natural frequencies of the bridge are 3.83, 15.32, 34.46, and 61.26 Hz, respectively.
To investigate the effect of vehicle and road roughness characteristics, two different scenarios are considered: TR-I and TR-II, each reflecting different traffic conditions and road surface characteristics.

The average mass of the vehicles is considered to be 1500 kg for TR-I and 500 kg for TR-II. In the spring-damper-mass model, the spring stiffness is assumed to be 170 kN/m, with a damping ratio of 20\% for all the vehicles. The arrival times for the vehicles is generated using Poison's distribution with an occurrence rate equal to 2. 
\begin{figure}
	\centering
		\includegraphics[scale=.70]{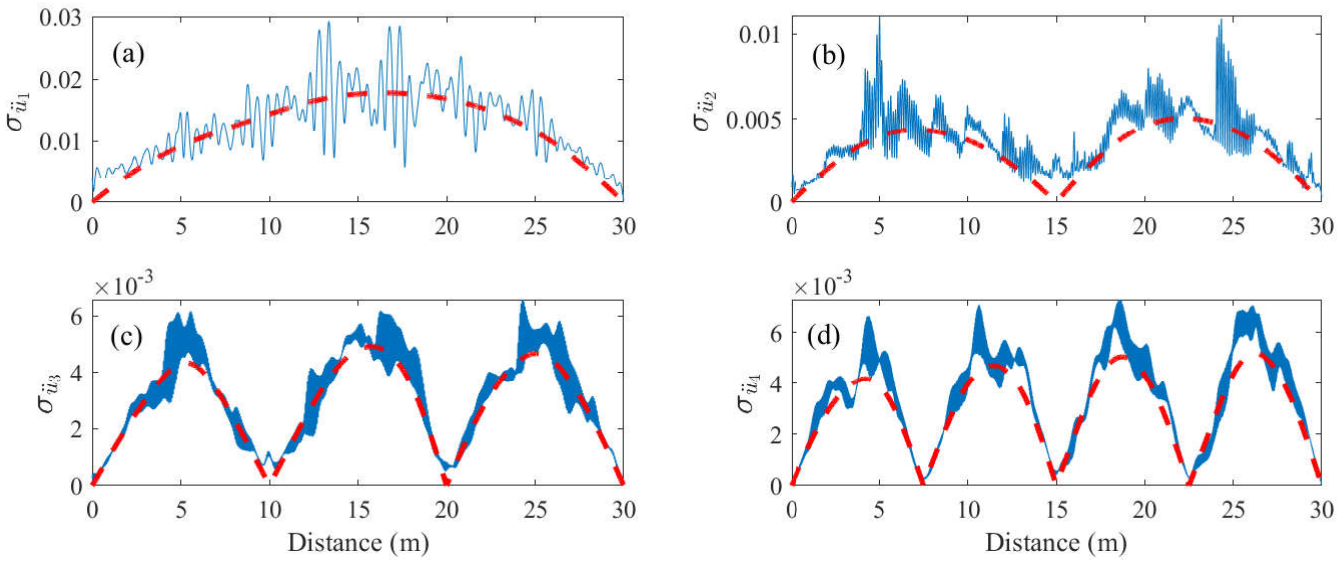}
	   \caption{Patterns of the SD of the decomposed response from the measured bridge responses for the first four modes in TR-I example}
	\label{C6FIG:msTR1}
\end{figure}
\begin{figure}
	\centering
		\includegraphics[scale=.70]{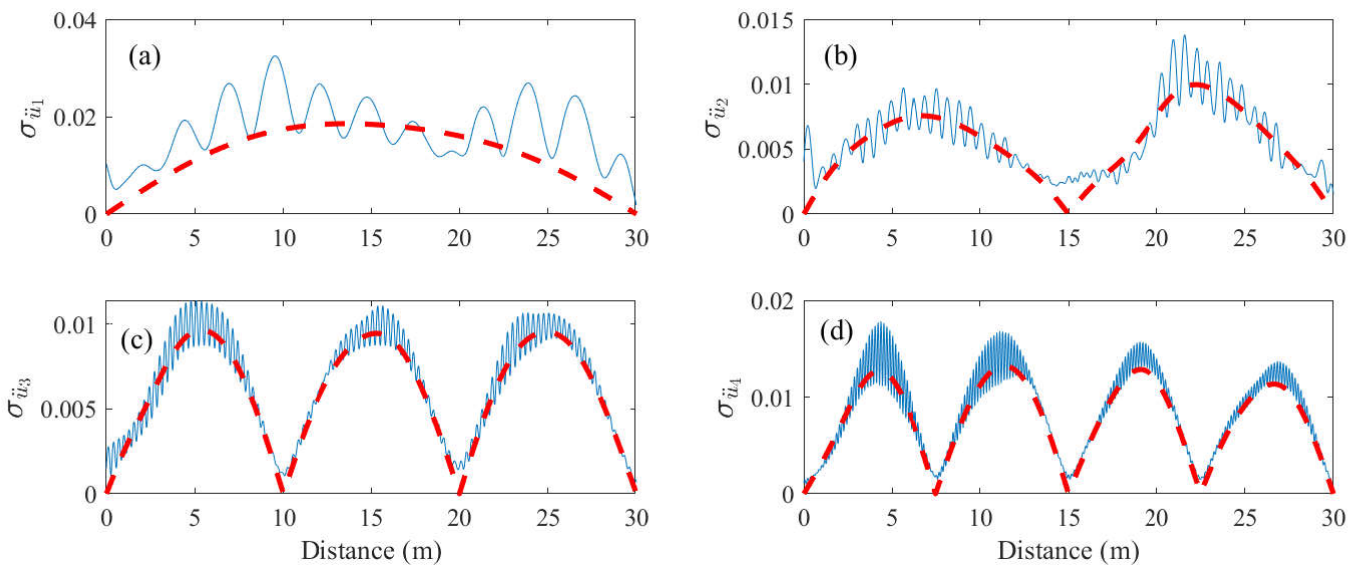}
	   \caption{Patterns of the SD of the decomposed response from the measured bridge responses for the first four modes in TR-II example}
	\label{C6FIG:msTR2}
\end{figure}

To simulate road roughness for TR-I and TR-II, $G_d(\kappa_0)$ is taken as $0.25\times 10^{-6}$ and $1\times 10^{-6}$, respectively. In TR-I, the mean velocity of the traffic, consisting of 25 vehicles~\footnote{Herein, responses are generated considering the fixed number of vehicles for the convenience of data storage and computation. Otherwise, response data can also be generated from a continuous stream of traffic.}, is considered as 20 m/s.
In contrast, the velocity of the instrumented vehicle is set at a relatively low value of 4 m/s.
To examine the effect of higher speed traffic in TR-II, the mean velocity of the traffic, consisting of 20 vehicles, is taken as 30 m/s. The velocity of the instrumented vehicle is considered to be 20 m/s. For the simulation, 50 runs of the instrumented vehicle are conducted in both scenarios, TR-I and TR-II. The velocity and mass of the vehicles, assumed to follow uniform distributions, are generated using the Latin hypercube sampling with variations of $\pm 2.5\%$ and $\pm 20\%$ around their mean values for the velocity and mass of the vehicles, respectively.
\par

\begin{table}
    \centering
\caption{Mean values of the identified natural frequencies and damping ratios in the VBI model, with corresponding standard deviations in parentheses }
\label{C6tab:wzeta_TR12}
    \begin{tabular}{c|cc|cc} \hline 
 \multirow{2}{1cm}{Mode} & \multicolumn{2}{c|}{TR-I}& \multicolumn{2}{c}{TR-II}\\ 
   & Natural frequency (Hz) & Damping ratio (in \%) &Natural frequency (Hz) &Damping ratio (in \%) \\
  \hline
  1 &  3.85 (0.23)&  1.88 (0.40\%)& 3.91 (0.14) &6.82 (1.01\%)\\
  2 &  15.27 (0.28)&  1.03 (0.30\%)& 15.36 (0.54) &1.55 (0.22\%)\\
  3& 34.45 (0.24)&  0.48 (0.08\%)& 34.60 (0.95) &0.69 (0.11\%)\\
  4& 60.38 (0.51)&  0.21 (0.05\%)& 60.33 (1.60)&0.41 (0.07\%)\\
     \hline
     \end{tabular}
\end{table}

The frequencies and damping ratios are identified using EFDD method, and the mean and standard deviation of these identified values are given in Table~\ref{C6tab:wzeta_TR12}.
To identify the mode shapes corresponding to these frequencies using the response from the moving sensor, the proposed statistical approach 
is used after decomposing the sensor responses into modal responses (i.e., $\ddot{u}_i$) using the EMD technique.

\textbf{Mode shape estimation using variance: }Following the decomposition of the responses for each simulation, the ensemble standard deviation (SD) of those decomposed modal responses are computed at each measuring point of the moving sensor. The SD of the modal responses ($\sigma_{\ddot{u}_i}$) corresponding to different frequencies are shown in Figs.~\ref{C6FIG:msTR1}, and~\ref{C6FIG:msTR2} for the scenarios TR-I and TR-II, respectively, with moving average values represented by thick dotted lines. 

\begin{figure}
	\centering
		\includegraphics[scale=.65]{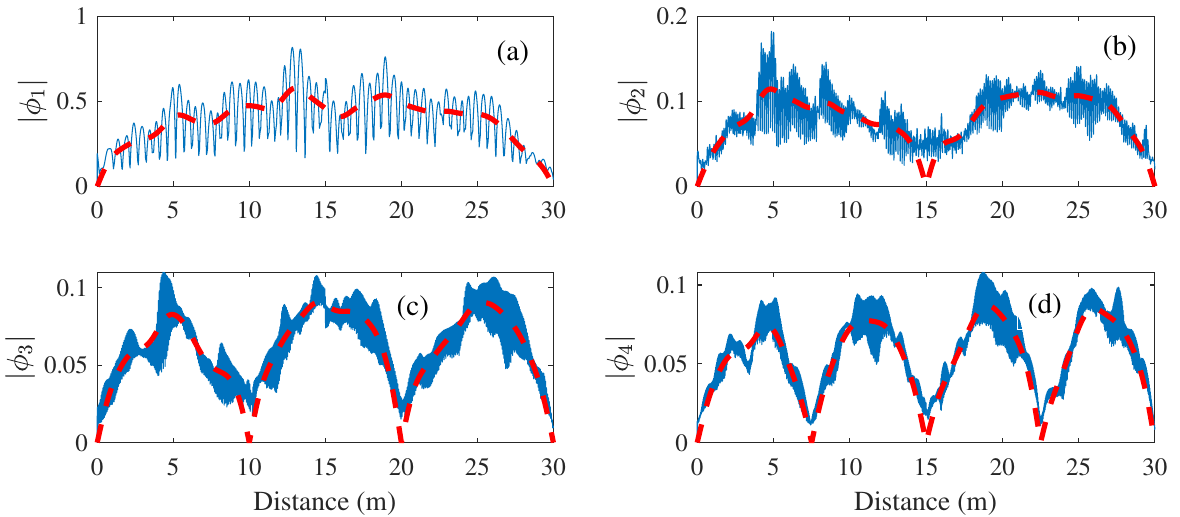}
	   \caption{Absolute value of mode shapes for the first four modes in (a-d), respectively, along the length of the bridge in TR-I example}
	\label{C6FIG:MS_EPS_TR1}
\end{figure}

\begin{figure}
	\centering
		\includegraphics[scale=.65]{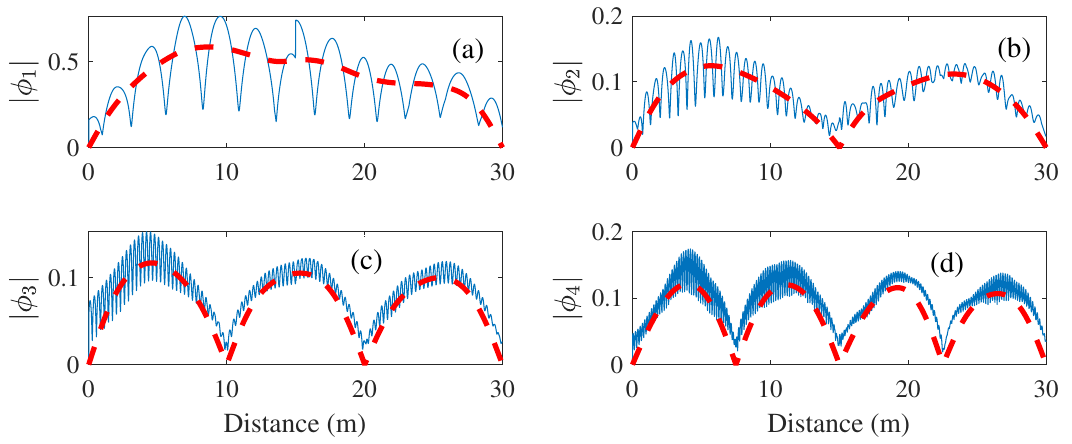}
	   \caption{Absolute value of mode shapes for the first four modes in (a-d), respectively, along the length of the bridge in TR-II example}
	\label{C6FIG:MS_EPS_TR2}
\end{figure}

\textbf{Mode shape estimation using EPS:} In this approach, the magnitude of the mode shapes is obtained using the EPS method, as given in "Mode shape estimation using evolutionary power spectrum" subsection. 
EPS is first calculated for each modal response. Rows of the EPS with spectral peaks represent squared mode shape values, from which the magnitudes are obtained by taking the square root. These are shown in Fig.~\ref{C6FIG:MS_EPS_TR1} for TR-I scenario, and Fig.~\ref{C6FIG:MS_EPS_TR2} for TR-II scenario, with thick dotted lines indicating the moving averages.


\begin{figure}
	\centering
		\includegraphics[scale=.75]{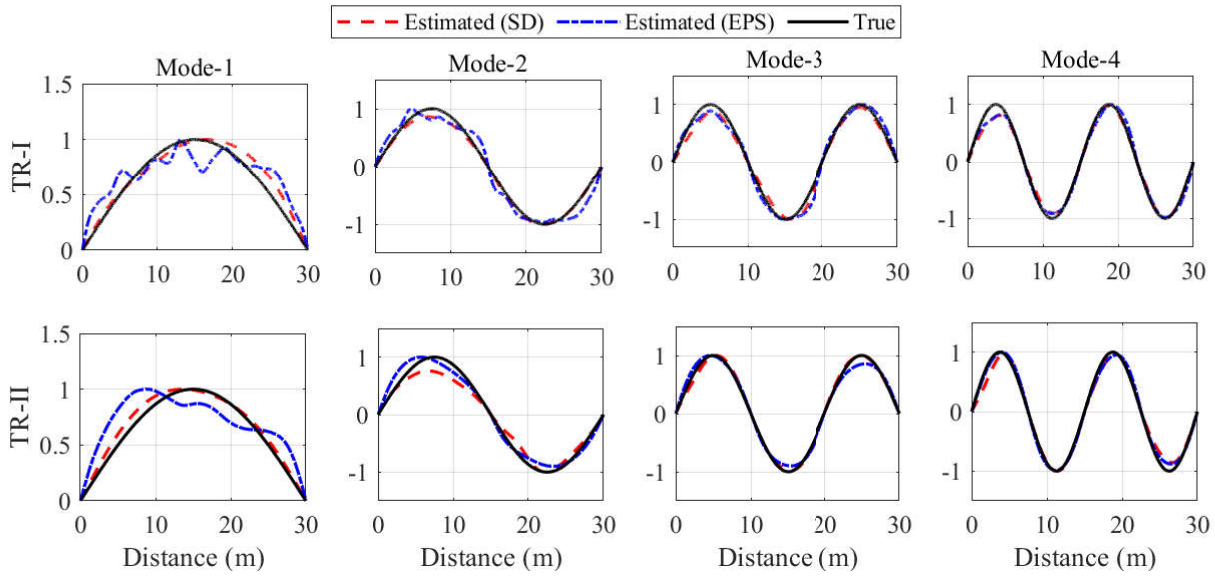}
	   \caption{Comparison of identified mode shapes (obtained using variances) for the first four modes for both TR-I and TR-II scenarios}
	\label{C6FIG:msTR-1_2}
\end{figure}

In order to identify the mode shape patterns, we can determine the signs of the mode shape components using engineering judgment. These identified mode shape patterns are then normalized to ensure a maximum value of one for comparison purpose. The resulting normalized mode shapes are compared to the true mode shapes of the simply supported bridge in Fig.~~\ref{C6FIG:msTR-1_2}. It is observed that the identified mode shapes obtained using the variance of the measurements (labeled as "Estimated (SD)" in the figure) are matching quite well with the true mode shapes for both TR-I and TR-II scenarios. However, in the TR-II scenario, the first mode shape obtained via the evolutionary power spectrum method ("Estimated (EPS)" in the figure) shows noticeable deviation from the true mode shape.  This discrepancy is likely due to the higher velocity of the moving sensor in the TR-II scenario, which results in fewer recorded data points, reducing the accuracy of the EPS-based mode shape estimation, where data from the time domain transferred to the frequency domain.

\section{Conclusion}
This paper presents a method for identifying the modal parameters of bridges, i.e., mode shapes, natural frequencies, and damping ratios, using a single moving sensor, so as to provide mode shapes with enhanced spatial resolution. It is shown that the moving sensor response contains the modal frequencies which can be obtained using any system identification technique; the EFDD technique is used in this paper to obtain the frequencies and damping ratios.
For mode shape identification, both known and unknown input scenarios are considered. In the case of known inputs, the mode shapes are expressed as a linear combination of appropriate Basic Orthonormal Polynomials (BOPs). The weights for these BOPs are estimated through a nonlinear least squares optimization technique. For unknown random traffic loading, it is shown that the variance of the second time derivative of the generalized coordinate, $\ddot{q}_i$ remains almost constant over the time span of the sensor traversing the bridge. Thus, the variance of the modal responses become proportional to the square of the corresponding mode shapes.
Leveraging this fact, and employing EMD to obtain the modal responses, the absolute value of the mode shape components are obtained as: (1) the standard deviation (SD) of the decomposed modal responses, or (2) from the evolutionary power spectrum of the decomposed modal responses.


Numerical examples using a simply supported bridge are used to demonstrate the mode shape identification methods. For the known input case, two types of inputs are considered: (a) Gaussian white noise (GWN) excitation at a specific location and (b) known moving masses with known velocities and arrival times (representing moving traffic).
It is observed that the mode shapes are identified quite accurately with high spatial resolution in both known input scenarios. However, for the known moving mass scenario, all the modes are excited and, hence, identified, while for the known excitation at a particular location, all the modes may not be excited and, therefore, may not be identified.
To validate the statistical approach for unknown random traffic loading, the same bridge is analyzed under traffic loads where vehicles are modeled as spring-damper-mass systems. The effects of vehicle speed and road roughness are examined through two numerical examples: one with low vehicle velocity and low-amplitude roughness, and another with high vehicle velocity and significant road roughness.
It is observed that the estimated mode shapes closely align with the "true" mode shapes when identified using both the time domain (using ensemble variance) and frequency domain (using evolutionary power spectrum) methods, particularly for cases involving low moving sensor velocity. However, the time domain approach performs better, especially in scenarios where the velocity of the moving sensor is high. 

The statistical method effectively bypasses the need to know the exact excitation force, making it highly practical for real-world applications. The results confirmed that even with a single moving sensor, it is possible to achieve reliable spatially exhaustive modal identification, offering a cost-efficient alternative to traditional dense sensor arrays. Future research may explore application and validation of this approach on actual bridge structures under varying environmental and traffic conditions.

\section{Acknowledgments}
Financial support for this work has been provided by the Science and Engineering Research Board (SERB, DST, India), under grant numbers ECR/2017/003430 \& SERB/CE/2023789. The financial support is gratefully acknowledged.

\section{Data Availability Statement}
Some or all data, models, or code that support the findings of this study are available from the corresponding author upon reasonable request.

\appendix
\section{Computation of BOPs}\label{C6append-1}
The displacement of a linear system can be represented as a linear combination of a set of basis functions, such as its mode shapes.
Suppose that the displacement $U(x,t)$ can be expressed by a Maclaurin series~\cite{hassanabadi2013new}:
\begin{equation}\label{C6eq:res_appen1}
    U(x,t) = q_1(t) P_1(x) + q_2(t) P_2(x) + \cdots + q_{n_0-m}(t) P_{n_0-m}(x); \quad \textrm{and  } n_0 = m+1, m+2, \cdots
\end{equation}
where $P_{n_0-m}(x)$, with $n_0$ controlling the number of terms retained for approximation, are called basic polynomials (BPs) which should satisfy the boundary conditions. For example, for a simply supported beam:
\begin{equation}
    U(x,t) = 0 \ \ \textrm{at }x=0,L \quad \textrm{and} \quad \frac{\partial^2 U(x,t)}{\partial x^2}=0  \ \ \textrm{at }x=0,L
\end{equation}
So the basic polynomials satisfying the boundary conditions can be calculated as~\cite{hassanabadi2013new}:
\begin{equation}
    P_i(x) = T_i(x)p_i + x^{m+i-1} \quad \textrm{where,  } T_i(x) = [1,\ x, \cdots, x^{m-1}]_{1\times m}\ \ \textrm{and $m=$ number of constraints} 
\end{equation}
where $p_i$ can be computed as:
\begin{equation}
    p_i = C^{-1}q_i
\end{equation}
The values of $C$ and $q_i$ for a simply supported beam are as follows ($m=4,\ x_0=0$ and $x_1=1$):
\begin{equation}
    C = 
    \begin{pmatrix}
        &1 &x_0 &x_0^2 & x_0^3 \\ &0 &0 &2 &6x_0 \\
        &1 &x_1 &x_1^2 & x_1^3 \\ &0 &0 &2 &6x_1 \\
    \end{pmatrix} 
    \quad \textrm{and} \quad 
    q_i=-
    \begin{pmatrix}
        &x_0^{m+i-1} \\ & (m+i-1)(m+i-2)x_0^{m+i-3} \\ &x_1^{m+i-1} \\ & (m+i-1)(m+i-2)x_1^{m+i-3} 
    \end{pmatrix}
\end{equation} 
Now, the BOPs can be extracted from BPs by orthogonalizing using a recursive Gram-Schmidt algorithm as:
\begin{equation}
\begin{matrix}
    \bar{P}_1(x) = \frac{P_1(x)}{<P_1(x),P_1(x)>^{1/2}} \\
    \bar{P}_i(x) = \frac{P_i(x)- \sum_{k=1}^{i-1} \bar{P}_k(x) <P_i(x),\bar{P}_k(x)>}{|| P_i(x) - \sum_{k=1}^{i-1} \bar{P}_k(x) <P_i(x),\bar{P}_k(x)> ||}
\end{matrix}
\quad \textrm{where }
\begin{cases}
    ||f(x)|| = <f(x),f(x)>^{1/2} \\
    <P_i(x),P_j(x)> = \int_{0}^{1} P_i(x) P_j(x) dx
\end{cases}
\end{equation}


\section{Time independence of variance of \lowercase{$\ddot{q}_n$}}\label{C6append-2}
In case the forces or masses arrive at the bridge at random times $t_k$, and with random velocity $v_k$, then the resulting force $f(x,t)$ can be expressed as~\cite{sniady1984vibration}:
\begin{equation}
    f(x,t) = \sum_{k=1}^{N_v(0,t)} A_k S(t-t_k)\delta \left[x-v_k(t-t_k)\right]
\end{equation}
Here, the amplitudes of the $k$th moving force or mass, $A_k$ are random variables assumed to be mutually independent and also independent of the time instants $t_k$. It is further assumed that $\mathsf{E}[A_k]=\mathsf{E}[A]=$constant, and $\mathsf{E}[A_k^2]=\mathsf{E}[A^2]=$constant.
The deterministic function $S(t-t_k)$ describes the variation of the forces with time, and specifically, for moving masses, $S(t-t_k)$ will be a unitary function during the period when a moving mass is present on the bridge. The velocities $v_k$ are considered random variables, independent of the instants $t_k$, and $N_v(0,t)$ represents a general counting stochastic process.

To demonstrate that the variances of $\ddot{q}_n(t)$ are time-independent, the solutions are formulated following the approach outlined in ~\cite{sniady1984vibration}.
The vibration response in modal co-ordinate is separated into the steady-state (ss, for $t\geq t_k$) and transient solutions (ts, for $t\geq t_k+T_k$) as:
\begin{equation}\label{C6eq:q_ss_ts}
\begin{matrix}
    q_n^{(ss)}(t-t_k,T_k) = \frac{1}{M_n} \int_{t_k}^{t} h_n(t-\tau)S(\tau-t_k) \phi_n \left[(L/T_k)(\tau-t_k)\right] d\tau
    \\
    q_n^{(ts)}(t-t_k-T_k,T_k) = \frac{1}{M_n} \int_{2t_k}^{t_k+T_k} h_n(t-\tau)S(\tau-t_k) \phi_n \left[(L/T_k)(\tau-t_k)\right] d\tau
    \end{matrix}
\end{equation}
where, $h_n(t-\tau)=\omega_{dn}^{-1} e^{-\zeta_n \omega_n (t-\tau)} sin (\omega_{dn}(t-\tau))$, and $T_k=L/v_k$ is the time taken to cross the bridge by $k$th force or moving mass. Evaluating the integrals in Eq.~\ref{C6eq:q_ss_ts}, we get:
\begin{equation}\label{C6eq:q_ss_ts_modi}
\begin{matrix}
    \begin{matrix}
        q_n^{(ss)}(t-\tau,T_k) =& \frac{2}{mL\Upsilon_n} \Bigg[ a_{1n} sin[\beta_n\{t-\tau\}] + a_{2n} cos[\beta_n\{t-\tau\}] \\
        &+ e^{-\zeta_n\omega_n\{t-\tau\}} \left( a_{3n} sin[w_{dn}\{t-\tau\}] - a_{2n} cos[w_{dn}\{t-\tau\}] \right) \Bigg] 
    \end{matrix}
    \\
    q_n^{(ts)}(t-\tau-T_k,T_k) = \frac{2}{mL\Upsilon_n}  e^{-\zeta_n\omega_n\{t-\tau-T_k\}} \left( b_{1n} sin[w_{dn}\{t-\tau-T_k\}] + b_{2n} cos[w_{dn}\{t-\tau-T_k\}] \right)
    \end{matrix}
\end{equation}
where the coefficients are defined as follows:
\begin{equation}
    \begin{matrix}
    \begin{matrix}
        \Upsilon_n = \left[\omega_n^2-\beta_n^2 \right]^2 + 4\zeta_n^2\omega_n^2 \beta_n^2; & \beta_n = n\pi/T_k; & a_{1n}= \omega_n^2-\beta_n^2
    \end{matrix}
    \\ 
    \begin{matrix}
        a_{2n} = -2\zeta_n \omega_n \beta_n; \qquad a_{3n} = \frac{\beta_n}{\omega_{dn}} \{ 2\zeta_n^2 \omega_n^2 - \left[\omega_n^2-\beta_n^2 \right] \}
    \end{matrix}
    \\
    \begin{matrix}
        b_{1n} = (-1)^n \omega_{dn}^{-1} \left[ a_{1n}\beta_n + a_{2n} \zeta_n \omega_n \right] + e^{-\zeta_n \omega_n T_k}\left[ a_{3n} cos(\omega_{dn}T_k) + a_{2n} sin(\omega_{dn}T_k) \right]
    \end{matrix}
     \\ 
     \begin{matrix}
        b_{2n} = (-1)^n a_{2n} + e^{-\zeta_n \omega_n T_k} \left[ a_{3n} sin(\omega_{dn}T_k) - a_{2n} cos(\omega_{dn}T_k) \right]
     \end{matrix}
    \end{matrix}
\end{equation}

From Eq.~\ref{C6eq:q_ss_ts_modi}, we can compute $\ddot{q}_n$ as follows:
\begin{equation}\label{C6eq:qn_ddot_ss_ts}
\begin{matrix}
\begin{matrix}
    \ddot{q}_n^{(ss)}(t-\tau,T_k) = & \frac{2}{mL\Upsilon_n} \Bigg[-\beta_n^2 \left( a_{1n} sin[\beta_n\{t-\tau\}] + a_{2n} cos[\beta_n\{t-\tau\}] \right) \\
     &+ \{\zeta_n^2 \omega_n^2 - \omega_{dn}^2 \} e^{-\zeta_n\omega_n\{t-\tau\}} \Bigl\{ a_{3n} sin[w_{dn}\{t-\tau\}] - a_{2n} cos[w_{dn}\{t-\tau\}] \Bigr\} \Bigg]    
\end{matrix} 
    \\
    \ddot{q}_n^{(ts)}(t-\tau-T_k,T_k) = \left(\zeta_n^2 \omega_n^2 - \omega_{dn}^2 \right) q_n^{(ts)}(t-\tau-T_k,T_k)
\end{matrix}
\end{equation}

In case of a stream of forces, the arrival of forces on the bridge span is modeled as a Poisson process with an arrival rate $\lambda_o$.
The probability density function can be represented by discrete sets as $\sum_{k=1}^{N_v}p_k \delta(T-T_k)$, where $p_k$ is the probability that the time of crossing the bridge $T$ is equal to $T_k$. This discrete probability framework for crossing times is discussed in detail in~\cite{sniady1984vibration}. The variance of $\ddot{q}_n(t)$ can then be computed as: 
\begin{equation}\label{C6eq:sig_ss_tI}
    \begin{matrix}
        \sigma_{\ddot{q}_n^{ss}}^2 (t) = \sigma_{\ddot{q}_n^{ss}}^2  = \frac{4\lambda_o \mathsf{E}[A^2]}{(mL)^2 } \sum_{k=1}^{N_v} \frac{p_k}{\Upsilon_n^2 } \Bigl[ \beta_n^4 \frac{T_k \Upsilon_n }{2} + \beta_n^2 \{\zeta_n^2 \omega_n^2 - \omega_{dn}^2 \} a_{1n} a_{3n} I_{1k}(\zeta_n, \omega_n,\beta_n,\omega_{dn}) 
        \\ \\
        - \beta_n^2 \{\zeta_n^2 \omega_n^2 - \omega_{dn}^2 \} a_{1n} a_{2n} I_{2k}(\zeta_n, \omega_n,\beta_n,\omega_{dn}) + \beta_n^2 \{\zeta_n^2 \omega_n^2 - \omega_{dn}^2 \} a_{2n} a_{3n} I_{2k}(\zeta_n, \omega_n,\omega_{dn},\beta_n)
        \\ \\
        - \beta_n^2 \{\zeta_n^2 \omega_n^2 - \omega_{dn}^2 \} a_{2n} a_{2n} I_{3k}(\zeta_n, \omega_n,\beta_n,\omega_{dn})
        \\ \\
        + \beta_n^2 \{\zeta_n^2 \omega_n^2 - \omega_{dn}^2 \} a_{3n} a_{1n} I_{1k}(\zeta_n, \omega_n,\omega_{dn},\beta_n) + \beta_n^2 \{\zeta_n^2 \omega_n^2 - \omega_{dn}^2 \} a_{3n} a_{2n} I_{2k}(\zeta_n, \omega_n,\omega_{dn},\beta_n)
        \\ \\
        + \{\zeta_n^2 \omega_n^2 - \omega_{dn}^2 \}^2 a_{3n} a_{3n} I_{1k}(2\zeta_n, \omega_n,\omega_{dn},\omega_{dn}) - \{\zeta_n^2 \omega_n^2 - \omega_{dn}^2 \}^2 a_{3n} a_{2n} I_{2k}(2\zeta_n, \omega_n,\omega_{dn},\omega_{dn})
        \\ \\
        - \beta_n^2 \{\zeta_n^2 \omega_n^2 - \omega_{dn}^2 \} a_{2n} a_{1n} I_{2k}(\zeta_n, \omega_n,\beta_n,\omega_{dn}) - \beta_n^2 \{\zeta_n^2 \omega_n^2 - \omega_{dn}^2 \} a_{2n} a_{2n} I_{3k}(\zeta_n, \omega_n,\omega_{dn},\beta_n)
        \\ \\
        - \{\zeta_n^2 \omega_n^2 - \omega_{dn}^2 \}^2 a_{2n} a_{3n} I_{2k}(2\zeta_n, \omega_n, \omega_{dn}, \omega_{dn}) + \{\zeta_n^2 \omega_n^2 - \omega_{dn}^2 \}^2 a_{2n} a_{2n} I_{3k}(2\zeta_n, \omega_n,\omega_{dn},\omega_{dn})
        \Bigl]
    \end{matrix}
\end{equation}
where,
\begin{equation}
    \begin{matrix}
        I_{1k}(a,b,c,d) = \int_{t-T_k}^t e^{-ab\{t-\tau\}} sin[c\{t-\tau\}] sin[d\{t-\tau\}] d\tau
        \\
        I_{2k}(a,b,c,d) = \int_{t-T_k}^t e^{-ab\{t-\tau\}} sin[c\{t-\tau\}] cos[d\{t-\tau\}] d\tau
        \\
        I_{3k}(a,b,c,d) = \int_{t-T_k}^t e^{-ab\{t-\tau\}} cos[c\{t-\tau\}] sin[d\{t-\tau\}] d\tau
    \end{matrix}
\end{equation}
It is evident from Eq.~\ref{C6eq:sig_ss_tI} that the function $\sigma_{\ddot{q}_n^{ss}} (t)$ is independent of time due to the stationarity of the arrival process.
As noted by Sniady~\cite{sniady1984vibration}, the time-independence of $\sigma_{{q}_n^{ts}}^2 (t)$ implies that $\sigma_{\ddot{q}_n^{ts}}^2 (t)$ is also time-independent, since ${\ddot{q}_n^{ts}} (t)$ is just a product of ${{q}_n^{ts}} (t)$ with a deterministic variable, as shown in Eq.~\ref{C6eq:qn_ddot_ss_ts}. Consequently, the variance of $\ddot{q}_n(t)$ is independent of time.


%
%
\bibliography{references}

\begin{thebibliography}{}

\bibitem[\protect\citeauthoryear{}{AlHamaydeh and Ghazal~Aswad}{2022}]{alhamaydeh2022structural}
AlHamaydeh, M. and Ghazal~Aswad, N. (2022).
\newblock ``Structural health monitoring techniques and technologies for large-scale structures: Challenges, limitations, and recommendations.''\ {\em Practice Periodical on Structural Design and Construction}, 27(3), 03122004.

\bibitem[\protect\citeauthoryear{}{Brewick and Smyth}{2015}]{brewick2015exploration}
Brewick, P.~T. and Smyth, A.~W. (2015).
\newblock ``Exploration of the impacts of driving frequencies on damping estimates.''\ {\em Journal of Engineering Mechanics}, 141(3), 04014130.

\bibitem[\protect\citeauthoryear{}{Brincker et~al.\@}{2001a}]{brincker2001damping}
Brincker, R., Ventura, C.~E., and Andersen, P. (2001a).
\newblock ``Damping estimation by frequency domain decomposition.''\ {\em Proceedings of IMAC 19: A Conference on Structural Dynamics: februar 5-8, 2001, Hyatt Orlando, Kissimmee, Florida, 2001}, Society for Experimental Mechanics,  698--703.

\bibitem[\protect\citeauthoryear{}{Brincker et~al.\@}{2001b}]{brincker2001modal}
Brincker, R., Zhang, L., and Andersen, P. (2001b).
\newblock ``Modal identification of output-only systems using frequency domain decomposition.''\ {\em Smart materials and structures}, 10(3), 441.

\bibitem[\protect\citeauthoryear{}{Cara et~al.\@}{2014}]{cara2014estimating}
Cara, F.~J., Juan, J., and Alarc{\'o}n, E. (2014).
\newblock ``Estimating the modal parameters from multiple measurement setups using a joint state space model.''\ {\em Mechanical Systems and Signal Processing}, 43(1-2), 171--191.

\bibitem[\protect\citeauthoryear{}{Cara~Ca{\~n}as et~al.\@}{2016}]{cara2016modal}
Cara~Ca{\~n}as, F.~J., Juan~Ruiz, J., and Alarc{\'o}n~{\'A}lvarez, E. (2016).
\newblock ``Modal identification of structures from roving input data by means of maximum likelihood estimation of the state space model.''\ {\em Proceedings of ISMA2016 International Conference on Noise and Vibration Engineering and USD2016 International Conference on Uncertainty in Structural Dynamics}, Industriales,  2923--2931.

\bibitem[\protect\citeauthoryear{}{Eshkevari et~al.\@}{2020}]{eshkevari2020bridge}
Eshkevari, S.~S., Matarazzo, T.~J., and Pakzad, S.~N. (2020).
\newblock ``Bridge modal identification using acceleration measurements within moving vehicles.''\ {\em Mechanical Systems and Signal Processing}, 141, 106733.

\bibitem[\protect\citeauthoryear{}{Gerlough et~al.\@}{1955}]{gerlough1955use}
Gerlough, D.~L., Schuhl, A., et~al.\@ (1955).
\newblock ``Use of poisson distribution in highway traffic. the probability theory applied to distribution of vehicles on two-lane highways.

\bibitem[\protect\citeauthoryear{}{Hassanabadi et~al.\@}{2013}]{hassanabadi2013new}
Hassanabadi, M.~E., Nikkhoo, A., Amiri, J.~V., and Mehri, B. (2013).
\newblock ``A new orthonormal polynomial series expansion method in vibration analysis of thin beams with non-uniform thickness.''\ {\em Applied Mathematical Modelling}, 37(18-19), 8543--8556.

\bibitem[\protect\citeauthoryear{}{Hosamo}{2012}]{hosamo2012study}
Hosamo, M. (2012).
\newblock ``A study of the source traffic generator using poisson distribution for abr service.''\ {\em Modelling and Simulation in Engineering}, 2012(1), 408395.

\bibitem[\protect\citeauthoryear{}{Huang et~al.\@}{1998}]{huang1998empirical}
Huang, N.~E., Shen, Z., Long, S.~R., Wu, M.~C., Shih, H.~H., Zheng, Q., Yen, N.-C., Tung, C.~C., and Liu, H.~H. (1998).
\newblock ``The empirical mode decomposition and the hilbert spectrum for nonlinear and non-stationary time series analysis.''\ {\em Proceedings of the Royal Society of London. Series A: mathematical, physical and engineering sciences}, 454(1971), 903--995.

\bibitem[\protect\citeauthoryear{}{ISO/TC et~al.\@}{1995}]{technical1995mechanical}
ISO/TC, T.~C., Vibration, M., Measurement, S. S.~S., of~Mechanical~Vibration, E., and as~Applied~to Machines, S. (1995).
\newblock {\em Mechanical Vibration--Road Surface Profiles--Reporting of Measured Data}, Vol. 8608.
\newblock International Organization for Standardization.

\bibitem[\protect\citeauthoryear{}{Iwankiewicz and {\'S}niady}{1984}]{iwankiewicz1984vibration}
Iwankiewicz, R. and {\'S}niady, P. (1984).
\newblock ``Vibration of a beam under a random stream of moving forces.''\ {\em Journal of Structural Mechanics}, 12(1), 13--26.

\bibitem[\protect\citeauthoryear{}{Jana et~al.\@}{2019}]{jana2019fisher}
Jana, D., Mukhopadhyay, S., and Ray-Chaudhuri, S. (2019).
\newblock ``Fisher information-based optimal input locations for modal identification.''\ {\em Journal of Sound and Vibration}, 459, 114833.

\bibitem[\protect\citeauthoryear{}{Jian et~al.\@}{2024}]{jian2024robotic}
Jian, X., Lai, Z., Bacsa, K., Fu, Y., Koh, C.~G., Sun, L., Wieser, A., and Chatzi, E. (2024).
\newblock ``A robotic automated solution for operational modal analysis of bridges with high-resolution mode shape recovery.''\ {\em Journal of Structural Engineering}, 150(8), 04024081.

\bibitem[\protect\citeauthoryear{}{Lin and Yang}{2005}]{lin2005use}
Lin, C. and Yang, Y. (2005).
\newblock ``Use of a passing vehicle to scan the fundamental bridge frequencies: An experimental verification.''\ {\em Engineering structures}, 27(13), 1865--1878.

\bibitem[\protect\citeauthoryear{}{Malekjafarian et~al.\@}{2022}]{malekjafarian2022review}
Malekjafarian, A., Corbally, R., and Gong, W. (2022).
\newblock ``A review of mobile sensing of bridges using moving vehicles: Progress to date, challenges and future trends.''\ {\em Structures}, Vol.~44, Elsevier,  1466--1489.

\bibitem[\protect\citeauthoryear{}{Malekjafarian and OBrien}{2014}]{malekjafarian2014identification}
Malekjafarian, A. and OBrien, E.~J. (2014).
\newblock ``Identification of bridge mode shapes using short time frequency domain decomposition of the responses measured in a passing vehicle.''\ {\em Engineering Structures}, 81, 386--397.

\bibitem[\protect\citeauthoryear{}{Matarazzo et~al.\@}{2022}]{matarazzo2022crowdsourcing}
Matarazzo, T.~J., Kondor, D., Milardo, S., Eshkevari, S.~S., Santi, P., Pakzad, S.~N., Buehler, M.~J., and Ratti, C. (2022).
\newblock ``Crowdsourcing bridge dynamic monitoring with smartphone vehicle trips.''\ {\em Communications engineering}, 1(1), 29.

\bibitem[\protect\citeauthoryear{}{MATLAB}{2023}]{MATLAB}
MATLAB (2023).
\newblock {\em version: 9.13.0 (R2023a)}.
\newblock The MathWorks Inc., Natick, Massachusetts 01760, USA, $<$https://www.mathworks.com$>$.

\bibitem[\protect\citeauthoryear{}{McGetrick et~al.\@}{2009}]{mcgetrick2009theoretical}
McGetrick, P.~J., Gonzlez, A., and OBrien, E.~J. (2009).
\newblock ``Theoretical investigation of the use of a moving vehicle to identify bridge dynamic parameters.''\ {\em Insight-Non-Destructive Testing and Condition Monitoring}, 51(8), 433--438.

\bibitem[\protect\citeauthoryear{}{Nayek et~al.\@}{2018}]{nayek2018mass}
Nayek, R., Mukhopadhyay, S., and Narasimhan, S. (2018).
\newblock ``Mass normalized mode shape identification of bridge structures using a single actuator-sensor pair.''\ {\em Structural Control and Health Monitoring}, 25(11), e2244.

\bibitem[\protect\citeauthoryear{}{Pasca et~al.\@}{2022}]{pasca2022pyoma}
Pasca, D.~P., Aloisio, A., Rosso, M.~M., and Sotiropoulos, S. (2022).
\newblock ``Pyoma and pyoma\_gui: a python module and software for operational modal analysis.''\ {\em SoftwareX}, 20, 101216.

\bibitem[\protect\citeauthoryear{}{{\'S}niady}{1984}]{sniady1984vibration}
{\'S}niady, P. (1984).
\newblock ``Vibration of a beam due to a random stream of moving forces with random velocity.''\ {\em Journal of Sound and Vibration}, 97(1), 23--33.

\bibitem[\protect\citeauthoryear{}{Tokognon et~al.\@}{2017}]{tokognon2017structural}
Tokognon, C.~A., Gao, B., Tian, G.~Y., and Yan, Y. (2017).
\newblock ``Structural health monitoring framework based on internet of things: A survey.''\ {\em IEEE Internet of Things Journal}, 4(3), 619--635.

\bibitem[\protect\citeauthoryear{}{Yang et~al.\@}{2025}]{yang2025indirect}
Yang, D., Yuan, Y., Zhang, J., and Au, F.~T. (2025).
\newblock ``Indirect bridge modal identification enhanced by iterative vehicle response demodulation.''\ {\em Mechanical Systems and Signal Processing}, 223, 111831.

\bibitem[\protect\citeauthoryear{}{Yang and Lee}{2018}]{yang2018damping}
Yang, J.~P. and Lee, W.-C. (2018).
\newblock ``Damping effect of a passing vehicle for indirectly measuring bridge frequencies by emd technique.''\ {\em International Journal of Structural Stability and Dynamics}, 18(01), 1850008.

\bibitem[\protect\citeauthoryear{}{Yang and Chang}{2009}]{yang2009extraction}
Yang, Y. and Chang, K. (2009).
\newblock ``Extraction of bridge frequencies from the dynamic response of a passing vehicle enhanced by the emd technique.''\ {\em Journal of sound and vibration}, 322(4-5), 718--739.

\bibitem[\protect\citeauthoryear{}{Yang et~al.\@}{2014}]{yang2014constructing}
Yang, Y., Li, Y., Chang, K.~C., et~al.\@ (2014).
\newblock ``Constructing the mode shapes of a bridge from a passing vehicle: a theoretical study.''\ {\em Smart Structures and Systems}, 13(5), 797--819.

\bibitem[\protect\citeauthoryear{}{Yang and Lin}{2005}]{yang2005vehicle}
Yang, Y. and Lin, C. (2005).
\newblock ``Vehicle--bridge interaction dynamics and potential applications.''\ {\em Journal of sound and vibration}, 284(1-2), 205--226.

\bibitem[\protect\citeauthoryear{}{Yang et~al.\@}{2024}]{yang2024bridge}
Yang, Y., Xu, W., Gao, A., Yang, Q., and Gao, Y. (2024).
\newblock ``Bridge damage identification based on synchronous statistical moment theory of vehicle--bridge interaction.''\ {\em Computer-Aided Civil and Infrastructure Engineering}.

\bibitem[\protect\citeauthoryear{}{Yang et~al.\@}{2004}]{yang2004extracting}
Yang, Y.-B., Lin, C., and Yau, J. (2004).
\newblock ``Extracting bridge frequencies from the dynamic response of a passing vehicle.''\ {\em Journal of Sound and Vibration}, 272(3-5), 471--493.

\bibitem[\protect\citeauthoryear{}{Yi et~al.\@}{2021}]{yi2021damage}
Yi, T.-H., Zhang, J., Qu, C.-X., and Li, H.-N. (2021).
\newblock ``Damage detection for decks of concrete girder bridges using the frequency obtained from an actively excited vehicle.''\ {\em Smart Structures and Systems, An International Journal}, 27(1), 101--114.

\bibitem[\protect\citeauthoryear{}{Zeng et~al.\@}{2023}]{zeng2023automation}
Zeng, J., Xie, Y.-L., Kim, Y.~H., and Wang, J. (2023).
\newblock ``Automation in bayesian operational modal analysis using clustering-based interpretation of stabilization diagram.''\ {\em Journal of Civil Structural Health Monitoring}, 13(2), 443--467.

\bibitem[\protect\citeauthoryear{}{Zhang et~al.\@}{2022}]{zhang2022detecting}
Zhang, J., Yi, T.-H., Qu, C.-X., and Li, H.-N. (2022).
\newblock ``Detecting hinge joint damage in hollow slab bridges using mode shapes extracted from vehicle response.''\ {\em Journal of Performance of Constructed Facilities}, 36(1), 04021109.

\bibitem[\protect\citeauthoryear{}{Zhu et~al.\@}{2012}]{zhu2012wireless}
Zhu, D., Guo, J., Cho, C., Wang, Y., and Lee, K.-M. (2012).
\newblock ``Wireless mobile sensor network for the system identification of a space frame bridge.''\ {\em Ieee/Asme Transactions On Mechatronics}, 17(3), 499--507.

\end{thebibliography}

\end{document}